\documentclass[twocolumn,aps,prx,nofootinbib,10pt]{revtex4-2}

\usepackage{amsmath}
\usepackage{amssymb}
\usepackage{wasysym}
\usepackage{graphicx}
\usepackage{hyperref}
\usepackage{dsfont}
\usepackage{newtxtext}
\usepackage[varvw]{newtxmath}
\usepackage{bm}
\usepackage[percent]{overpic}
\hypersetup{
    colorlinks,
    linkcolor={blue},
    citecolor={blue},
    urlcolor={blue}
}

\newcommand{\rme}{\mathrm{e}}
\newcommand{\rmi}{\mathrm{i}}

\newcommand{\Tc}{{T_{\mathrm{c}}}}

\usepackage{xcolor}
\usepackage[normalem]{ulem}

\begin{document}

%%%%%%%%%%%%%%%%%%%%%%%%%%%%%%%%%%%%%%%%%%%%%%%%%%%%%%%%%%%%%%%%%%%%%%%
\title{Higher-Rank Spin Liquids and Spin Nematics from Competing Orders in Pyrochlore Magnets}
%%%%%%%%%%%%%%%%%%%%%%%%%%%%%%%%%%%%%%%%%%%%%%%%%%%%%%%%%%%%%%%%%%%%%%%

\author{Niccol\`o Francini}
\author{Lukas Janssen}
\author{Daniel Lozano-G\'omez}

\affiliation{Institut f\"ur Theoretische Physik and W\"urzburg-Dresden Cluster of Excellence ct.qmat, TU Dresden, 01062 Dresden, Germany}

%%%%%%%%%%%%%%%%%%%%%%%%%%%%%%%%%%%%%%%%%%%%%%%%%%%%%%%%%%%%%%%%%%%%%%%
\begin{abstract}
Pyrochlore magnets have proven to provide an excellent arena for the realization of a variety of many-body phenomena such as classical and quantum order-by-disorder, as well as spin liquid phases described by emergent gauge field theories. These phenomena arise from the competition between different symmetry-breaking magnetic orders. In this work, we consider a subspace of the most general bilinear nearest-neighbor Hamiltonian on the pyrochlore lattice, parameterized by the local interaction parameter $J_{z\pm}$, where three symmetry-breaking phases converge. We demonstrate that for small values of $|J_{z\pm}|$, a conventional $\mathbf q=0$ ordered phase is selected by a thermal order-by-disorder mechanism. For $|J_{z\pm}|$ above a certain finite threshold, a novel spin-nematic phase is stabilized at low temperatures. Instead of the usual Bragg peaks, the spin-nematic phase features lines of high intensity in the spin structure factor. At intermediate temperatures above the low-temperature orders, a rank-2 U(1) classical spin liquid is realized for all $J_{z\pm} \neq 0$. We fully characterize all phases using classical Monte-Carlo simulations and a self-consistent Gaussian approximation.
\end{abstract}
%%%%%%%%%%%%%%%%%%%%%%%%%%%%%%%%%%%%%%%%%%%%%%%%%%%%%%%%%%%%%%%%%%%%%%%

%%%%%%%%%%%%%%%%%%%%%%%%%%%%%%%%%%%%%%%%%%%%%%%%%%%%%%%%%%%%%%%%%%%%%%%
\date{February 19, 2025}
%%% REPLACE \today -> [actual submission date] BEFORE ARXIV SUBMISSION
%%%%%%%%%%%%%%%%%%%%%%%%%%%%%%%%%%%%%%%%%%%%%%%%%%%%%%%%%%%%%%%%%%%%%%%

\maketitle

%%%%%%%%%%%%%%%%%%%%%%%%%%%%%%%%%%%%%%%%%%%%%%%%%%%%%%%%%%%%%%%%%%%%%%%
\section{Introduction}
\label{sec:intro}
%%%%%%%%%%%%%%%%%%%%%%%%%%%%%%%%%%%%%%%%%%%%%%%%%%%%%%%%%%%%%%%%%%%%%%%

In the search for novel states of matter and associated exotic many-body phenomena, the study of frustrated magnets plays a pivotal role.
Traditionally, magnetic frustration is categorized into two distinct classes. \emph{Geometric frustration} occurs when the lattice geometry prevents spins from simultaneously satisfying all interaction constraints~\cite{Springer_frust,Balents2010,savary2012}. This phenomenon is commonly observed in lattices with corner-sharing triangular motifs, such as the Kagome~\cite{kagome_PhysRevB.71.024401} lattice in two dimensions and the pyrochlore~\cite{Rau2018FrustratedQR} lattice in three dimensions. \emph{Exchange frustration}, on the other hand, arises when the interactions between spins are inherently competing, independent of the lattice geometry. Recent research in this context has focused on magnets with highly anisotropic interactions, induced by significant spin-orbit coupling~\cite{trebst22}.
A particularly intriguing scenario arises when both geometric and exchange frustration mechanisms are simultaneously relevant. This situation can occur in pyrochlore compounds of the form $A_2M_2$O$_7$, where $A$ and $M$ are rare-earth and transition-metal elements, respectively~\cite{Rau2018FrustratedQR}. The coexistence of both types of frustration can lead to the emergence of novel exotic phenomena that are absent when only one frustration mechanism is present.

The pyrochlore lattice, composed of a network of corner-sharing tetrahedra, provides an excellent platform for studying novel states of matter and associated many-body phenomena. This lattice has enabled significant research into spin liquids~\cite{Castelnovo2008,Yan2020_rank2_u1,Moessner1998_low_temp,Taillefumier2017_xxz,Benton2016_Pinch_line,Benton2012_seeing,EvansPhysRevX.12.021015,lozano-gomez2023,chung_gingras_2023arxiv,lozano_2024arxiv}, order-by-disorder (ObD) phenomena~\cite{villain1980,belorizky1980,zhitomirsky2012,zhitomirsky2013,savary2012,chern10,Noculak_HDM_2023,Hickey_2024arxiv}, magnetic fragmentation~\cite{Brooks-Bartlett2014,Petit2016,Benton2016_quantum_origins}, and topological magnons~\cite{Onose2010}.
Of particular interest is the realization of spin liquids, disordered yet strongly-correlated phases of matter, whose low-temperature behaviors are described by emergent gauge field theories~\cite{Yan_2024_PRB_classification}. In the case of classical spin liquids, these phases are characterized by an extensive ground state degeneracy, reflected in the occurrence of low-energy flat bands in the spectrum of the interaction matrix~\cite{Chung_PRL,Yan_2024_PRB_classification}. These exotic phases of matter have turned out to be quite elusive and are usually associated with a delicate and balanced competition between different magnetic orders, preventing the formation of a symmetry-broken state at low temperatures. Recent work has illustrated how such delicate balance might be realized at points in the interaction parameter space where multiple magnetic orders intersect~\cite{Taillefumier2017_xxz,lozano-gomez2023}. The stability of a spin liquid at these points may be however hampered by an ObD mechanism, whereas thermal or quantum fluctuations select a set of states out of the spin liquid's extensively degenerate manifold resulting in a symmetry-breaking spin configuration. Indeed, previous studies~\cite{Noculak_HDM_2023} have demonstrated how such an ObD selection takes place at phase boundaries between distinct magnetic phases, typically resulting in a magnetically ordered $\textbf{q}=0$ phase at low temperatures.

\begin{figure}[b!]
    \centering
    \begin{overpic}[width=1.\columnwidth]{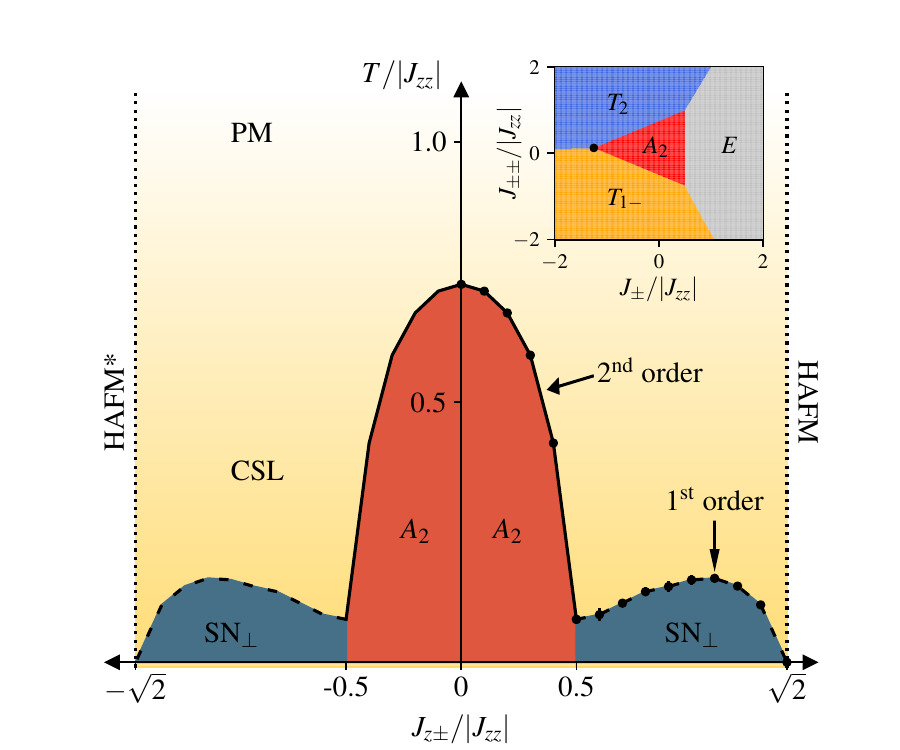}

    \end{overpic}
    \caption{Finite-temperature phase diagram of the model in Eq.~\eqref{eq:general_bilinear_hamiltonian} as function of the coupling $J_{z\pm}$, with $J_{\pm}$ and $J_{\pm\pm}$ tuned to the triple point between the $A_2$, $T_{1-}$, and $T_2$ phases in the classical limit. As the temperature decreases, we observe a crossover from a high-temperature paramagnetic (PM) phase to an intermediate-temperature classical spin liquid (CSL), followed by a phase transition into one of two possible symmetry-broken states: a $\mathbf q=0$  all-in-all-out antiferromagnet ($A_2$) or a non-magnetic spin nematic~($\rm SN_\perp$) consisting of fluctuating components of the $T_{1-}$ and $T_2$ irreps. The black dots represent the transition temperatures obtained via classical Monte-Carlo simulations. Full (dashed) lines indicate continuous (discontinuous) phase transitions.
    The inset displays the zero-temperature phase diagram as a function of $J_{\pm}$ and $J_{\pm\pm}$ for fixed $J_{z\pm} > 0$ and $J_{zz} < 0$~\cite{Rau2018FrustratedQR}. The black circle marks the triple phase boundary investigated in this work. At this point, the ground state is extensively degenerate, as highlighted by the thick yellow line at $T=0$ in the main panel.}
    \label{fig:Schematic_phase_diagram_small}
\end{figure}

In this work, we present a detailed study of the distinct phases resulting from a thermal ObD selection for a classical spin model where
three
distinct long-range ordered phases intersect. We demonstrate that the strong competition between these phases results in a spin liquid regime at intermediate temperatures, characterized by twofold and fourfold pinch points in the spin structure factor. We show how the spin liquid phase is destabilized at low temperatures, resulting in a symmetry-breaking transition into a conventional $\mathbf q=0$ magnetically ordered phase or a novel spin-nematic ($\rm SN_\perp $) phase. We then show that the novel $\rm SN_\perp $ phase can be further stabilized at low temperatures across an extended region in parameter space, where only two distinct long-range ordered phases intersect.

We consider the bilinear nearest-neighbor Hamiltonian~\cite{Rau2018FrustratedQR}
\begin{eqnarray}
\mathcal{H}&=&\sum_{\langle ij \rangle} \Big\{ J_{zz} S_i^z S_j^z -J_{\pm}( S_i^+ S_j^- + S_i^- S_j^+) + J_{\pm\pm } (\gamma_{ij}S_i^+ S_j^+ +\rm{h.c.})\nonumber\\
&& \qquad +\ J_{z\pm}(\zeta_{ij}S_i^z( S_j^++S_j^-) +\rm{h.c.})\Big\},
\label{eq:general_bilinear_hamiltonian}
\end{eqnarray}
where $S^z_i$ and $S_i^\pm = S_i^x \pm \rmi S_i^y$ represent the components of the (pseudo) spin-$1/2$ on site $i$ in a \emph{local} coordinate frame, and the bond-dependent phase factors $\gamma_{ij} =  - \zeta_{ij}^*$ are given in Appendix~A of Ref.~\cite{Ross2011quantum_excitations}. In this work, we consider the classical limit of the Hamiltonian in Eq.~\eqref{eq:general_bilinear_hamiltonian} and mainly study the temperature phase diagram along
% first focus on
a line in parameter space parameterized by the coupling $J_{z\pm}$, which we dub the $A_2\oplus T_1\oplus T_2$ line, where two $\mathbf q=0$ antiferromagnetic phases, the $A_2$~\cite{sadeghiSpinHamiltonianOrder2015} and $T_2$~\cite{PC2000PRB} phases, meet with a $\mathbf q=0$ ferromagnetic phase, the $T_{1-}$~\cite{Thompson2017} phase. The competition between these three phases results in an extensively-degenerated ground state manifold which, at intermediate temperatures, leads to a rank-1 U(1) spin liquid phase for $J_{z\pm}=0$ and a rank-2 U(1) spin liquid for $J_{z\pm}\neq 0$. At low temperatures, the spin liquid phases are destabilized by symmetry-breaking ObD selection  to a $\mathbf q=0$ state for $|J_{z\pm}|/ |J_{zz}| \lesssim 0.5$ and to a spin-nematic phase for $0.5 \lesssim |J_{z\pm}|/ |J_{zz}|<\sqrt{2}$. Our results are summarized in the schematic phase diagram shown in Fig.~\ref{fig:Schematic_phase_diagram_small}.
We then demonstrate that an equivalent spin nematic state is stabilized at low temperatures on an entire plane in parameter space corresponding to the boundary between two conventionally ordered phases. This plane, referred to as the $T_1\oplus T_2$ plane, is where the ferromagnetic phase $T_{1-}$ intersects with only one of the antiferromagnetic phases, the $T_2$.

The remainder of the paper is organized as follows: In Sec.~\ref{sec:irreps}, we introduce the irreducible representations of the single-tetrahedron Hamiltonian and provide the parametrization describing the $A_2\oplus T_1\oplus T_2$ line. In Sec.~\ref{section:SCGA}, we briefly introduce the self-consistent Gaussian approximation and present the low-energy bands of the models parametrized by the $A_2\oplus T_1\oplus T_2$ line. Section~\ref{section:cMC} contains details of our Monte-Carlo approach and a list of thermodynamics quantities measured in the classical simulations. A detailed analysis for $J_{z\pm}=0$, i.e., the non-Kramers Hamiltonian, for which a rank-1 spin liquid phase is realized at intermediate temperatures, is provided in Sec.~\ref{sec:non-Kramers}. In Sec.~\ref{sec:Kramers}, we consider the case $|J_{z\pm}|>0$, for which a rank-2 U(1) spin liquid is realized at intermediate temperatures. In this section, we also discuss in detail the two distinct parameter regimes where $\mathbf q=0$ and $\rm SN_\perp $ phases, respectively, are selected at low temperatures. We conclude in Sec.~\ref{sec:discussion_conclusion}. Technical details are deferred to four appendices.

%%%%%%%%%%%%%%%%%%%%%%%%%%%%%%%%%%%%%%%%%%%%%%%%%%%%%%%%%%%%%%%%%%%%%%%
\section{Irreducible-representation analysis} \label{sec:irreps}
%%%%%%%%%%%%%%%%%%%%%%%%%%%%%%%%%%%%%%%%%%%%%%%%%%%%%%%%%%%%%%%%%%%%%%%

In this section, we present an overview of the irreducible-representation analysis conducted for a generic nearest-neighbor dipolar Hamiltonian on the pyrochlore lattice. In the classical limit, and exploiting the corner-sharing geometry of the pyrochlore lattice, the Hamiltonian in Eq.~\eqref{eq:general_bilinear_hamiltonian} can be written as
\begin{eqnarray}
    \mathcal{H}=\sum_\boxtimes \mathcal{H}_\boxtimes,
\end{eqnarray}
where $\sum_{\boxtimes}$ denotes the sum over all tetrahedra and $\mathcal{H}_\boxtimes$ represents the single-tetrahedron Hamiltonian. This single-tetrahedron Hamiltonian can be expressed in terms of irreducible representations (irreps) of the single tetrahedron group $T_d$~\cite{Yan2017_theory_multiphase,Wong2013},%
\footnote{An irreducible-representation analysis of this group identifies five irreps, namely the $A_2$, $E$, $T_2$, $T_{1-}$, where $A$ refers to one-dimensional irreps, $E$ refers to two-dimensional irreps, and $T$ refers to three-dimensional irreps~\cite{KTC_2024_phase,Wong2013,Yan2017_theory_multiphase}.}
\begin{eqnarray}
    \mathcal{H}_\boxtimes   &=& \frac{1}{2}\left[ a_{A_2} \mathrm{m}_{A_2}^2 + a_{E} \mathbf{m}_{E}^2 + a_{T_2} \mathbf{m}_{T_2}^2 \right. \nonumber\\
   && \left. +\  a_{T_{1-}} \mathbf{m}_{T_{1 -}}^2
  +  a_{T_{1+}} \mathbf{m}_{ T_{1 +}}^2 \right], \label{eq:irrep_decomp}
\end{eqnarray}
where the irreps $A_2$, $E$, $T_2$, $T_{1-}$, and $T_{1+}$ label the spin modes describing the all-in-all-out (AIAO)~\cite{sadeghiSpinHamiltonianOrder2015,Petit2016}, $\Gamma_5$~\cite{McClarty_2009,zhitomirsky2012,Sarkis2020YbGeO}, Palmer-Chalker~\cite{PC2000PRB,yahne2021_reentrance}, and two types of splayed ferromagnetic orders~\cite{Thompson2017,Ross2011quantum_excitations},%
\footnote{There are special cases for which the two $T_1$ irreps represent a colinear ferromagnet and a coplanar antiferromagnet.}
respectively. $\mathbf{m}_{I}$ is the spin mode in a single tetrahedron corresponding to the $I$th irrep, and $a_I$ is a function of the interaction parameters $\{J_{zz},J_{\pm},J_{\pm\pm},J_{z\pm}\}$ representing the internal energy cost associated with the $I$th irrep. For more details regarding the explicit form of the $\mathbf{m}_I$ spin modes and their corresponding single-tetrahedron energies $a_I$ we refer to Appendix~\ref{appendix:irreps}.

Recent work~\cite{Yan2017_theory_multiphase,Wong2013} has demonstrated the utility of the above irrep decomposition for generating ground state phase diagrams. In particular, this approach allows one to identify the irreps with the minimum energy cost $a_I$, and aids in predicting effective gauge theories describing spin liquid phases in cases when multiple irreducible representations are energetically degenerate and no symmetry-breaking transition occurs~\cite{Yan_2024_PRB_classification}.
In the context of the present work, we study the different phases accessed as a function of temperature when the $A_2$, $T_{1-}$, and the $T_2$ irreps compete, and the two remaining irreps (namely the $E$ and the $T_{1+}$ irreps) are gapped at higher energies. In terms of the interaction parameters in Eq.~\eqref{eq:general_bilinear_hamiltonian},
the region in parameter space where the $A_{2}$, $T_{1-}$, and $T_2$ irreps are degenerate  is defined by
the parametrization
\begin{eqnarray}
J_\pm=-\frac{J_{z\pm}^2}{J_{zz}}+\frac{3}{2}J_{zz}\quad \mathrm{and}\quad
    J_{\pm\pm}=-\frac{J_{z\pm}^2}{2J_{zz}} \label{eq:parametrization_A2_T1_T2}
\end{eqnarray}
with the constraints
\begin{eqnarray} \label{eq:constraints_A2_T1_T2}
    J_{zz}<0\quad \mathrm{and}\quad
    |J_{z\pm}|<\sqrt{2}|J_{zz}|.
\end{eqnarray}
The above functional form of the interaction parameters identifies a line in the interaction parameter space as a function of the coupling $J_{z\pm}$, which we refer to as the $A_2\oplus T_1\oplus T_2 $ line. The boundaries of this line are the points $J_{z\pm}=\pm \sqrt{2}|J_{zz}|$, which correspond to two spin liquid points, the Heisenberg antiferromagnet for $J_{z\pm}>0$ and its dual for $J_{z\pm}<0$~\cite{atlas_2024}. Furthermore, this parametrization can be segmented in two families of Hamiltonians: the non-Kramers Hamiltonian for which $J_{z\pm}=0$, and the Kramers Hamiltonian for which $J_{z\pm}\neq 0$~\cite{Rau2018FrustratedQR}.

In what follows, we map out the finite-temperature phase diagram as function of $J_{z\pm}$ along the $A_2\oplus T_1\oplus T_2 $ line, first focusing on the non-Kramers Hamiltonian in Sec.~\ref{sec:non-Kramers}, and then proceeding to the Kramers Hamiltonian in Sec.~\ref{sec:Kramers}.
The $A_2\oplus T_1\oplus T_2$ line is a subset of the $T_1\oplus T_2$ plane. As a consequence, a complete characterization of this subset will facilitate a comprehensive understanding of the entire $T_1\oplus T_2$ plane, which will be detailed towards the end of Sec.~\ref{sec:Kramers}. Before continuing, we note that the model in Eq.~\eqref{eq:general_bilinear_hamiltonian} features a duality under changing the sign of $J_{z\pm}$ followed by a $\pi/2$ rotation about the local $z$ axis of all spin~\cite{Rau2018FrustratedQR}. Consequently, for the Kramers Hamiltonian, we only present results for the case for which $J_{z\pm}\geq 0$, and extend our results to the $J_{z\pm}<0$ case using the duality relation. From this point on, and based on the parametrization of the $A_2\oplus T_1\oplus T_2 $ line, we assume units in which the Boltzmann constant $k_{\rm B}=1$ and measure all energies in units of $|J_{zz}|=1$.

%%%%%%%%%%%%%%%%%%%%%%%%%%%%%%%%%%%%%%%%%%%%%%%%%%%%%%%%%%%%%%%%%%%%%%%
\section{Self-Consistent Gaussian Approximation}
\label{section:SCGA}
%%%%%%%%%%%%%%%%%%%%%%%%%%%%%%%%%%%%%%%%%%%%%%%%%%%%%%%%%%%%%%%%%%%%%%%

The self-consistent Gaussian approximation (SCGA)~\cite{SCGA_Canals_kagome,SCGA_Canals_checker,SCGA_Canals_pyrochlore,Isakov2004,Conlon2010,Chung_PRL} is a classical method in which the hard spin length constraint, $\mathbf{S}_i^2=S^2$, is replaced by a soft constraint, $\sum_i \mathbf{S}_i^2 = N S^2$, where $N$ corresponds to the total number of sites on the lattice. The latter can be easily implemented by the cost of a single Lagrange multiplier $\lambda$. This approximation yields a quadratic Hamiltonian that can be solved exactly, and from which the spin correlation functions and effective low-temperature theories can be studied. The starting point of this analysis is to rewrite the Hamiltonian in Eq.~\eqref{eq:general_bilinear_hamiltonian} as
\begin{equation}
\mathcal{H}=\frac{1}{2}\sum_{i,j}\mathbf{S}_{i}^\top \, \mathbf{ M}_{ij}\,  \mathbf{S}_{j},
\label{eq:SCGA_general_Hamiltonian}
\end{equation}
where $\mathbf{M}_{ij}$ denotes the $3\times 3$ real-space interaction matrix for given fixed lattice sites $i$ and $j$. In the SCGA, we have direct access to the static spin structure factor, given as
\begin{equation}
    S({\mathbf{q}}) =\sum_{\mu,\nu} \sum_{\alpha} ( \beta\ \mathrm{M}^{\alpha\alpha}_{\mu\nu}(\mathbf{q})+\lambda)^{-1},\label{eq:SCGA-Sq}
\end{equation}
where $\beta$ is the inverse temperature, $\mathbf{M} (\mathbf{q})$ is the $12\times 12$ interaction matrix in reciprocal space, $\alpha$ labels the Cartesian spin components, and $\mu$ and $\nu$ label the sublattice structure. Note that the spin structure factor involves taking a trace with respect to the spin components $\alpha$, so Eq.~\eqref{eq:SCGA-Sq} includes only the diagonal spin components of the interaction matrix. %

\begin{figure}[tb!]
    \centering
    \begin{overpic}[width=0.8\columnwidth]{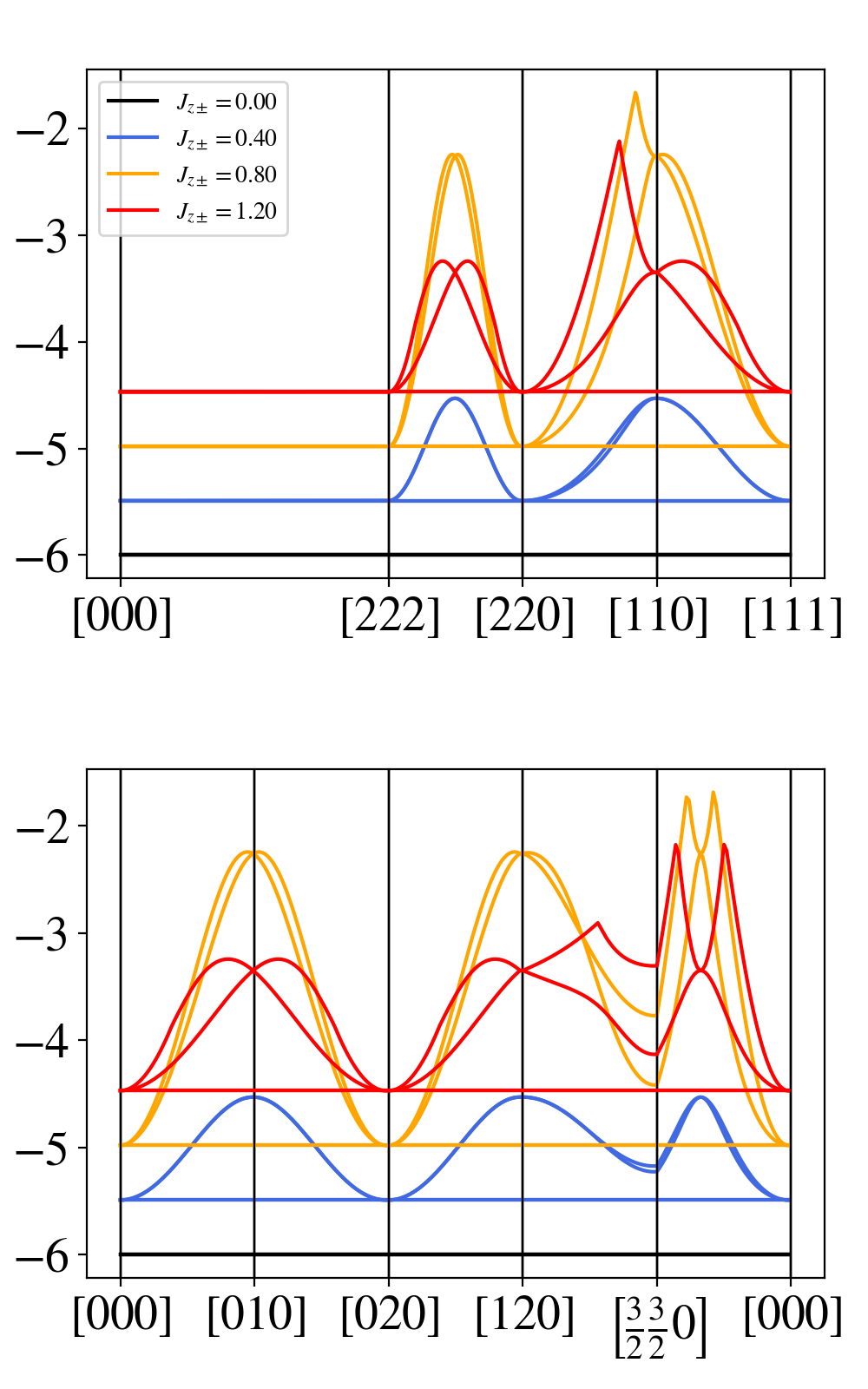}
    \put(5,97){(a)}
    \put(5,47){(b)}
    \put(-4,75){\rotatebox{90}{$E(\mathbf{q})$}}
    \put(-4,25){\rotatebox{90}{$E(\mathbf{q})$}}
    \end{overpic}
    \caption{Four lowest-energy bands in the spectrum of the interaction matrix $\mathbf{M}_{ij}$ for different fixed values of $J_{z\pm}$ on the $A_2 \oplus T_1 \oplus T_2 $ line, i.e., for $J_\pm$ and $J_{\pm\pm}$ chosen according to Eq.~\eqref{eq:parametrization_A2_T1_T2},
    along high-symmetry paths in
    (a)~the $[hh\ell]$ plane and
    (b)~the $[hk0]$ plane
    in reciprocal space. Different colors indicate different values of $J_{z\pm}$, given in the inset of panel~(a). For better visibility, different-colored bands have been offset by a constant energy shift of $0.5$ from black to red.}
\label{fig:bands}
\end{figure}
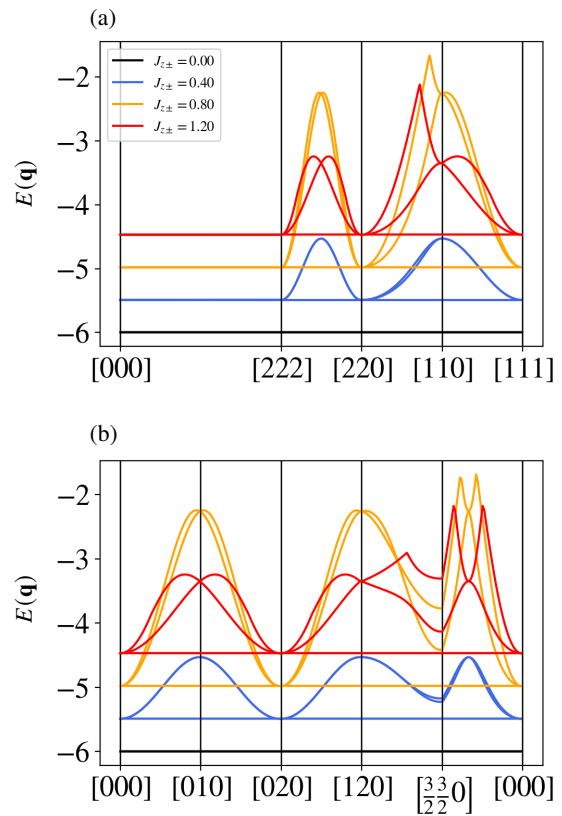

For the $A_2\oplus T_1\oplus T_2 $ line, the diagonalization of the interaction matrix $\mathbf{M} (\mathbf{q})$  results in the identification of four low-energy flat bands for $J_{z\pm}=0$ and two low-energy flat-bands for $J_{z\pm}\neq 0$, see Fig.~\ref{fig:bands}. At the level of the SCGA, the observation of low-energy flat bands in the spectrum of the interaction matrix indicates the realization of a stable spin liquid phase down to the lowest temperatures~\cite{Isakov2004,lozano-gomez2023}. The SCGA, however, does not account for higher-order terms in a free-energy expansion, which may result in the ObD selection of a symmetry-breaking phase at low temperatures. Consequently, in order to assess the validity of the approximation, complementary methods must be used to determine the ultimate fate of the systems in the zero-temperature limit.

%%%%%%%%%%%%%%%%%%%%%%%%%%%%%%%%%%%%%%%%%%%%%%%%%%%%%%%%%%%%%%%%%%%%%%%
\section{Classical Monte-Carlo simulations}
\label{section:cMC}
%%%%%%%%%%%%%%%%%%%%%%%%%%%%%%%%%%%%%%%%%%%%%%%%%%%%%%%%%%%%%%%%%%%%%%%

%%%%%%%%%%%%%%%%%%%%%%%%%%%%%%%%%%%%%%%%%%%%%%%%%%%%%%%%%%%%%%%%%%%%%%%
\subsection{Algorithmic details}
%%%%%%%%%%%%%%%%%%%%%%%%%%%%%%%%%%%%%%%%%%%%%%%%%%%%%%%%%%%%%%%%%%%%%%%

As an unbiased complementary method, we study the model in Eq.~\eqref{eq:general_bilinear_hamiltonian} using large-scale Monte-Carlo simulations based on the Metropolis \cite{metropolis1953} and overrelaxation algorithm \cite{creutz1987}, equipped with parallel tempering exchange \cite{marinari92,hukushima96,hukushima99}. In the simulations, the spins are treated as classical vectors of unit length $\mathbf{S}_i^2=1$.
For each lattice sweep of a Metropolis update, we combine 20 overrelaxation steps and propose a replica exchange of the configurations at adjacent temperatures. The procedure is repeated 20 times between two consecutive measurements on the lattice.
We consider pyrochlore lattices with $L^3$ tetrahedra and $N=4L^3$ sites, using $L=8, 10, 12$ and employing periodic boundary conditions.
The accumulated statistics is of the order $\mathcal{O}(10^5)$ for each point in parameter space~\cite{nhr-alliance}.

%%%%%%%%%%%%%%%%%%%%%%%%%%%%%%%%%%%%%%%%%%%%%%%%%%%%%%%%%%%%%%%%%%%%%%%
\subsection{Observables}
%%%%%%%%%%%%%%%%%%%%%%%%%%%%%%%%%%%%%%%%%%%%%%%%%%%%%%%%%%%%%%%%%%%%%%%

In our simulations, we measure the following thermodynamic observables:

\paragraph{Energy density.} The internal energy density $\varepsilon$ is defined by the energy per site,
\begin{equation}
    \label{eq:energy_density}
    \varepsilon \coloneqq  \frac{\langle\mathcal{H}\rangle}{N},
\end{equation}
where $N$ corresponds to the total number of sites.

\paragraph{Specific heat.} From the fluctuations of the energy, we measure the specific heat as
\begin{equation}
    \label{eq:specific_heat-MC}
    C \coloneqq \frac{\langle \mathcal{H}^2 \rangle - \langle \mathcal{H}\rangle^2}{NT^2},
\end{equation}
where $T$ is the absolute temperature. Alternatively, the specific heat can be computed from the energy density as $C = \partial \varepsilon / \partial T$.

\paragraph{Magnetic order parameters.} Magnetic order parameters can be formulated using the single-tetrahedron irreps $\mathbf{m}_{I}$, $I = A_2, E, T_2, T_{1\pm}$ as
\begin{equation}
    \label{eq:irreps_measures}
    \mathrm{m}_I = \left\langle \left| \frac{1}{L^3} \sum_{\boxtimes} \mathbf{m}_I \right| \right\rangle,
\end{equation}
where $\sum_{\boxtimes}$ denotes the sum over all tetrahedra on the lattice. The single-tetrahedron irreps $\mathbf{m}_{I}$ are given explicitly in terms of the spins in Appendix~\ref{appendix:irreps}.

\paragraph{Spin-nematic order parameter.} The single-site spin-nematic order parameter is defined as~\cite{Nematic_order_Shanon2006PRL,Nematic_order_Shanon2015PRB,Taillefumier2017_xxz,francini24}
\begin{equation}
\label{eq:single_site_nematic_op}
    Q_\perp^{\rm site}= \left\langle \left| \frac{1}{N} \sum_{i} \begin{pmatrix}
        (S_{i}^x)^2 - (S_{i}^y)^2 \\ \\
        2S_{i}^x S_{i}^y
    \end{pmatrix}   \right| \right\rangle ,
\end{equation}
and is sensitive to the local $xy$ arrangement of the spins.

\paragraph{Static spin structure factor.} We also measure the static spin structure factor defined as
\begin{equation}
    \label{eq:spin_sructure_factor}
    S({\mathbf{q}}) = \frac{1}{N} \sum_{i,j} \langle \mathbf{S}_i \cdot \mathbf{S}_j \rangle \rme^{- \rmi\mathbf{q}\cdot(\mathbf{R}_i - \mathbf{R}_j)},
\end{equation}
where $\mathbf{R}_i$ is the position vector of the site $i$ on the lattice. The spin structure factor allows for a direct comparison between our Monte-Carlo results and SCGA predictions. Moreover, in a spin liquid phase, it clearly reflects the properties of the underlying emergent gauge field theory. Therefore, in the main text, we focus on the spin structure factor rather than neutron structure factors, which are more directly accessible experimentally. Results for the neutron structure factors at representative points along the $A_2\oplus T_1 \oplus T_2$ line are presented in Appendix~\ref{appendix:Structure_factors}.

%%%%%%%%%%%%%%%%%%%%%%%%%%%%%%%%%%%%%%%%%%%%%%%%%%%%%%%%%%%%%%%%%%%%%%%
\section{Non-Kramers case \texorpdfstring{$J_{z\pm}=0$}{J=0}}
\label{sec:non-Kramers}
%%%%%%%%%%%%%%%%%%%%%%%%%%%%%%%%%%%%%%%%%%%%%%%%%%%%%%%%%%%%%%%%%%%%%%%

We begin our analysis of the $A_2\oplus T_1\oplus T_2$ line by first considering the point along this line at which $J_{z\pm}=0$. This scenario results in a non-Kramers Hamiltonian, with the local $z$ spin degrees of freedom transforming as magnetic dipoles and the local $xy$ degrees of freedom transforming as magnetic quadrupoles~\cite{Rau2018FrustratedQR,Petit2016PRBPrZr}.
In terms of irreps, the decoupling between the local $z$ and local $xy$ degrees of freedom is evident in the specific form of the $T_1$ irreps. These irreps correspond to two distinct splayed ferromagnetic configurations, where spins are either aligned along their local $z$ axis or confined to their local $xy$ plane~\cite{Rau2018FrustratedQR,Taillefumier2017_xxz}.
For the case $J_{z\pm}=0$ in the parametrization of the $A_2\oplus T_1\oplus T_2 $ line, the $T_{1-}$ corresponds to local-$xy$ splayed ferromagnet.

%%%%%%%%%%%%%%%%%%%%%%%%%%%%%%%%%%%%%%%%%%%%%%%%%%%%%%%%%%%%%%%%%%%%%%%
\subsection{SCGA results}
%%%%%%%%%%%%%%%%%%%%%%%%%%%%%%%%%%%%%%%%%%%%%%%%%%%%%%%%%%%%%%%%%%%%%%%

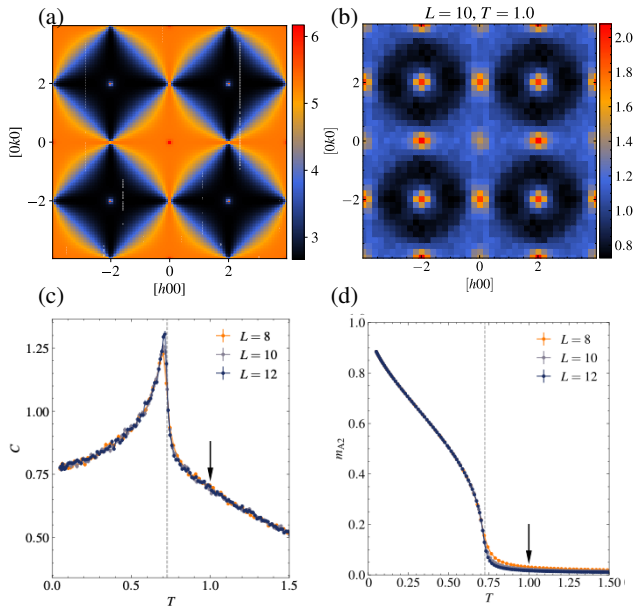
\begin{figure}[tb!]
    \centering
    \begin{overpic}[width=1.\columnwidth]{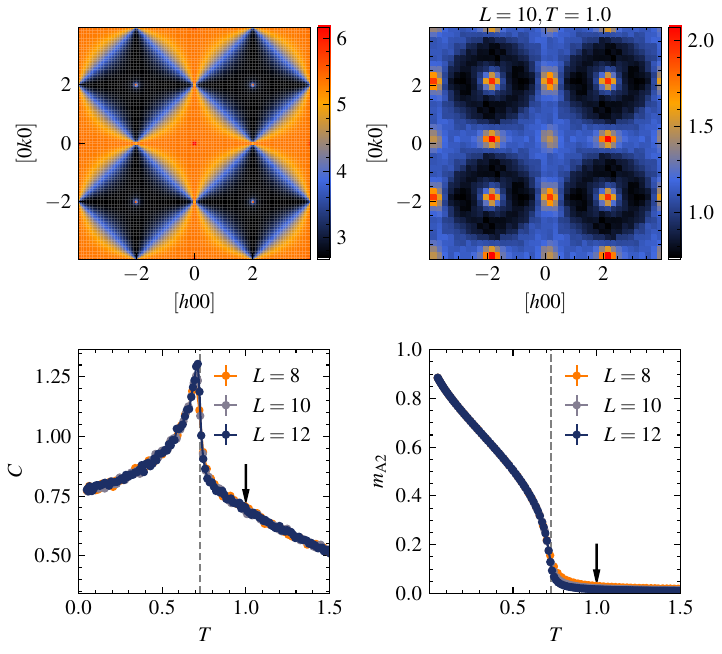}
    \put(5,88){(a)}
    \put(5,45){(c)}
    \put(55,88){(b)}
    \put(55,45){(d)}
    \end{overpic}
    \caption{%
    (a)~Static spin structure factor in the $[hk0]$ plane from SCGA in the low-temperature limit for the non-Kramers case $J_{z\pm} = 0$, describing the correlations in the rank-1 classical spin liquid.
    (b)~Same as~(a), but from classical Monte-Carlo simulations on an $L=10$ lattice for a fixed temperature $T = 1$ above $\Tc$.
    (c)~Specific heat as function of temperature $T$ for different lattice sizes, revealing a continuous transition in the 3D Ising universality class at $\Tc = 0.7266(3)$.
    (d)~Same as~(c), but for the $A_2$ order parameter $m_{A_2}$.
    Dashed lines in (c) and (d) indicate the critical temperature. Arrows indicate the temperature chosen in (b).
    }
\label{fig:non-kramers}
\end{figure}

As was previously stated, the observation of four low-energy flat bands in the spectrum of the interaction matrix would imply that, at the level of SCGA, a spin liquid phase is realized. Following the identification of the low-energy flat bands, we use the irrep fields, as defined in Appendix~\ref{appendix:irreps}, and the effective Hamiltonian obtained through an SCGA analysis to obtain an effective model describing the low-energy physics. This procedure is fully analogous to those performed in recent studies~\cite{Benton2016_Pinch_line,Yan2020_rank2_u1,lozano_2024arxiv}, and we refer the reader to these works for further details. The spin liquid phase predicted by the SCGA analysis results in the identification of two rank-1 gauge fields, namely
\begin{eqnarray}
      \bm{E}^\mathsf{A} &=& (2  \mathrm{m}_{T_{1-}}^x , -    \mathrm{m}_{T_{1-}}^y -\sqrt{3}\mathrm{m}_{T_{2}}^y, -   \mathrm{m}_{T_{1-}}^z + \sqrt{3}\mathrm{m}_{T_{2}}^z), \label{eq:gauge_field_A}\\
  \bm{E}^\mathsf{B} &=& (2 \mathrm{m}_{T_{2}}^x , + \sqrt{3}    \mathrm{m}_{T_{1-}}^y - \mathrm{m}_{T_{2}}^y, -    \sqrt{3}   \mathrm{m}_{T_{1-}}^z -   \mathrm{m}_{T_{2}}^z),\label{eq:gauge_field_B}
\end{eqnarray}
which fulfill energetically-imposed Gauss's law at low temperatures
\begin{equation}
  \partial_i  {E}^\mathsf{A}_i = 0 \quad \text{and}\quad
  \partial_i  {E}^\mathsf{B}_i = 0,\label{eq:Gauss_laws}
\end{equation}
where we have assumed summation over repeated indices. In the above expressions, the $\mathrm{m}_I^\alpha$ labels the $\alpha$ component of the $I$th irrep mode. The energetically-enforced Gauss's laws result in the observation of sharp twofold pinch features in the SCGA spin structure factor, see Fig.~\ref{fig:non-kramers}(a). We note that the gauge fields defined in Eqs.~\eqref{eq:gauge_field_A} and~\eqref{eq:gauge_field_B} do not have any component of the $A_2$ irrep. The absence of this contribution at low temperatures is due to the $A_2$ field adhering to its own Gauss's law, specifically $\partial_i \mathrm{m}_{A_2}=0$~\cite{atlas_2024,Benton2014_thesis}. Given that the $\mathrm{m}_{A_2} $ is a one-dimensional field, the Gauss's law implies that at low temperatures, it can only obtain a uniform constant value throughout the lattice. In the absence of long-range order, the fulfilment of this Gauss's law would imply that $\mathrm{m}_{A_2}=0$.
The effective gauge theory describing the spin liquid regime encompasses a total of six degrees of freedom, corresponding to the six components of the $T_{1-}$ and $T_2$ irreps, associated with the gauge fields $\mathbf{E}^\mathsf{A}$ and $\mathbf{E}^\mathsf{B}$. These are subject to two constraints provided by their respective Gauss's laws. Consequently, this framework identifies four flat bands for the $J_{z\pm}=0$ spin liquid, resulting from the subtraction of the number of constraints from the total degrees of freedom, in agreement with the results shown in Fig.~\ref{fig:bands}.

%%%%%%%%%%%%%%%%%%%%%%%%%%%%%%%%%%%%%%%%%%%%%%%%%%%%%%%%%%%%%%%%%%%%%%%
\subsection{Classical Monte-Carlo results}
%%%%%%%%%%%%%%%%%%%%%%%%%%%%%%%%%%%%%%%%%%%%%%%%%%%%%%%%%%%%%%%%%%%%%%%

To assess the stability of the spin liquid phase predicted by the SCGA analysis, we compare its predictions with the results obtained from classical Monte-Carlo simulations. For not too low temperatures, both approaches yield consistent results. In particular, at intermediate temperatures around $T \sim |J_{zz}| = 1$, the system effectively behaves as the classical spin liquid predicted by SCGA. The spin structure factors at $T=1$, see Fig.~\ref{fig:non-kramers}(b), resemble the SCGA prediction modulo thermal broadening of the pinch points, where the broadening of these features is associated with a non-zero population of gauge charges and therefore the violation of the Gauss's laws in Eq.~\eqref{eq:Gauss_laws}. In the low-temperature limit, the classical spin liquid regime is not stable: the system eventually reaches a critical temperature $\Tc$, below which a long-range $\mathbf q=0$ order develops due to a thermal ObD mechanism, which we discuss in further detail below..
At the transition, the specific heat shows a pronounced peak, as seen in Fig.~\ref{fig:non-kramers}(c).
Among the three competing long-range ordered phases at the $A_2\oplus T_1\oplus T_2 $ point, only one is selected, namely, the $A_2$ phase, as suggested by the corresponding magnetic order parameter $\mathrm m_{A_2}$, shown in Fig.~\ref{fig:non-kramers}(d). Moreover, we have explicitly verified that the spin structure factor develops the Bragg-peak pattern expected for $A_2$ order (not shown)~\cite{sadeghiSpinHamiltonianOrder2015}.
The AIAO state spontaneously breaks time-reversal symmetry, suggesting that the transition, if continuous, belongs to the 3D Ising universality class. Indeed, through a standard finite-size scaling analysis of the Binder cumulant associated with $\mathrm m_{A_2}$ (not shown), we find the correlation-length exponent $\nu = 0.636(5)$. Similarly, exploiting a finite-size scaling of the specific heat and the susceptibility associated with $\mathrm m_{A_2}$, we obtain the critical indices $\alpha=0.11(1)$ and $\gamma=1.19(2)$. These indices are consistent with a continuous 3D Ising transition~\cite{chang2024}.

The thermal ObD selection can be understood using classical low-temperature expansion, which yields an effective quadratic Hamiltonian for the spin fluctuations about a ground-state configuration~\cite{Noculak_HDM_2023}.  Through this expansion, we identify the spin configuration with the lowest energy fluctuations, which is then expected to be thermally selected at low temperatures~\cite{Yan2017_theory_multiphase}. For a $\mathbf{q}=0$ state, this expansion identifies 8 modes in reciprocal space.
We study the energy fluctuations about the three competing  $\mathbf{q}=0$ magnetic orders along the
$A_2\oplus T_1\oplus T_2$ line, where these three magnetic long-range orders compete.
The effective Hamiltonian obtained for fluctuations about the $T_{1-}$ and the $T_2$ states yields 8 quadratic dispersive bands.
In contrast, for the $A_2$ state, this expansion identifies four zero-energy flat bands and four dispersive high-energy bands.
As a consequence, the $A_2$ state is equipped with quadratic and at-least-quartic modes, resulting in lower energy fluctuations and a higher associated entropic weight~\cite{Noculak_HDM_2023}. The classical low-temperature expansion therefore suggests that the thermal ObD generates an entropic selection of the $A_2$ ground state. For more details on the expansion, we refer to Appendix~\ref{appendix:CLTE}.

The quadratic and at-least-quartic modes about the $A_2$ state leave a signature in the low-temperature value of the specific heat. In particular, if we assume that the zero-energy fluctuation modes in the classical low-temperature expansion
are quartic, from the equipartition theorem, the specific heat in units of $k_\mathrm{B}$ in the limit of low temperature follows,
\begin{equation} \label{eq:specific-heat-CLTE}
    C=\frac{1}{4}\left( \frac{n_2}{2} + \frac{n_4}{4} \right),
\end{equation}
where $n_2$ and $n_4$ are the numbers of quadratic and quartic modes, respectively, with $n_2+n_4=8$ being the total number of modes.
For the $A_2$ state, we then have $n_2 = n_4 = 4$, from which we obtain the prediction $C=3/4$ for the specific heat at zero temperature, which agrees with the numerical results shown in Fig.~\ref{fig:non-kramers}(c), confirming that the zero-energy modes in the effective quadratic theory are quartic modes.
%

%%%%%%%%%%%%%%%%%%%%%%%%%%%%%%%%%%%%%%%%%%%%%%%%%%%%%%%%%%%%%%%%%%%%%%%
\section{Kramers case \texorpdfstring{$J_{z\pm}\neq0$}{J=0}}\label{sec:Kramers}
%%%%%%%%%%%%%%%%%%%%%%%%%%%%%%%%%%%%%%%%%%%%%%%%%%%%%%%%%%%%%%%%%%%%%%%

%%%%%%%%%%%%%%%%%%%%%%%%%%%%%%%%%%%%%%%%%%%%%%%%%%%%%%%%%%%%%%%%%%%%%%%
\subsection{SCGA results}
%%%%%%%%%%%%%%%%%%%%%%%%%%%%%%%%%%%%%%%%%%%%%%%%%%%%%%%%%%%%%%%%%%%%%%%

We now discuss the Kramers case for which $J_{z\pm}\neq0$. As for the non-Kramers case, we first start by providing an SCGA analysis and then supplement this with a classical Monte-Carlo study. Similarly to the $J_{z\pm}=0$ case, the Hamiltonian describing the $A_2\oplus T_1\oplus T_2$ line for $J_{z\pm}\neq 0$ presents flat bands in the interaction matrix spectrum. In this case however, the number of flat bands reduces to two out of twelve. Nevertheless, in an SCGA analysis, the observation of low-energy flat bands implies the realization of
a spin liquid phase. In this case however, the system is described by three emergent gauge fields, of which two are rank-1 gauge fields, namely
\begin{align}
  \bm{E}^{A_{z\pm}} =& (2 \sin\phi\ \mathrm{m}_{T_{1-}}^x , -    \sin\phi \mathrm{m}_{T_{1-}}^y -\sqrt{3}\mathrm{m}_{T_{2}}^y,\nonumber\\& -   \sin\phi\ \mathrm{m}_{T_{1-}}^z + \sqrt{3}\mathrm{m}_{T_{2}}^z),\\
  \bm{E}^{B_{z\pm}} =& (2 \mathrm{m}_{T_{2}}^x , + \sqrt{3}    \sin\phi\ \mathrm{m}_{T_{1-}}^y - \mathrm{m}_{T_{2}}^y,\nonumber\\
  &-    \sqrt{3}   \sin\phi\ \mathrm{m}_{T_{1-}}^z -   \mathrm{m}_{T_{2}}^z),
\end{align}
and the remaining gauge field is a rank-2 field,
\begin{equation}
\begin{split}
      \mathbb{E}^{A_2+T_{1}+T_2} =& -\sin\phi\begin{bmatrix}
	\mathrm{m}_{A_2} & 0   &   0  \\
	0  & \mathrm{m}_{A_2}&  0 \\
	  &0 & \mathrm{m}_{A_2}
\end{bmatrix}\\
&+\frac{\cos\phi}{2}\begin{bmatrix}
	0 & \sqrt{3} \mathrm{m}_{T_2}^z    &    - \sqrt{3} \mathrm{m}_{T_2}^y   \\
	- \sqrt{3} \mathrm{m}_{T_2}^z  & 0&  \sqrt{3} \mathrm{m}_{T_2}^x \\
	 \sqrt{3} \mathrm{m}_{T_2}^y  &-\sqrt{3} \mathrm{m}_{T_2}^x& 0
\end{bmatrix}\\& +
2\sin\phi\cos\phi
\begin{bmatrix}
	0 & \mathrm{m}_{T_{1-}}^z &     \mathrm{m}_{T_{1-}}^y \\
	 \mathrm{m}_{T_{1-}}^z &0&   \mathrm{m}_{T_{1-}}^x \\
	 \mathrm{m}_{T_{1-}}^y &   \mathrm{m}_{T_{1-}}^x & 0
\end{bmatrix},
\end{split}
\end{equation}
where the angle $\phi$ corresponds to the decoupling angle between the two $T_1$ irreps, see Appendix~\ref{appendix:irreps}. At low energies, these fields are constraint by three energetically-imposed Gauss's law constraints
\begin{eqnarray}
    \partial_i {E}^{A_{z\pm}}_i &=& 0,\nonumber\\
     \partial_i {E}^{B_{z\pm}}_i &=& 0,\label{eq:gauss_laws_jzpm}\\
	\partial_i \mathbb{E}^{A_2+T_{1}+T_2}_{ij} &=& 0.\nonumber
\end{eqnarray}
Note that the construction of the rank-2 tensor $\mathbb{E}^{A_2+T_{1}+T_2}$ includes irrep fields that also appear in the construction of the rank-1 fields, $\bm{E}^{A_{z\pm}}$ and $ \bm{E}^{B_{z\pm}}$. Consequently, the Gauss's law constraints on the rank-1 fields can be seen as additional constraints on the rank-2 field. According to the Gauss's law constraints in Eq.~\eqref{eq:gauss_laws_jzpm} for the rank-2 field, the system realizes a rank-2 U(1) spin liquid characterized by fourfold pinch points in the spin structure factor~\cite{Davier2023}. The observation of these fourfold features is associated with the realization of fractons, excitations whose diffusion is severely restricted by the presence of charge, dipole, and possibly other multipole conservation laws~\cite{Davier2023,Pretko-2017,Pretko_fractonsPhysRevB.98.115134}. We emphasize that the observation of fourfold pinch points, either broad or sharp, is expected for all non-vanishing $J_{z\pm}$ on the $A_2\oplus T_1\oplus T_2$ line, as the general structure of the Gauss's laws and fields remains unchanged at the level of the SCGA.
The higher-rank spin liquid observed for $J_{z\pm} \neq 0$ is characterized by seven degrees of freedom, corresponding to the seven fluctuating fields (the components of the $A_2$, $T_{1-}$, and $T_2$ fields), and five constraints provided by the two scalar and three vector components of the Gauss's laws in Eq.~\eqref{eq:gauss_laws_jzpm}. As a result, this spin liquid is characterized by two low-energy flat bands, in agreement with the results shown in Fig.~\ref{fig:bands}.

%%%%%%%%%%%%%%%%%%%%%%%%%%%%%%%%%%%%%%%%%%%%%%%%%%%%%%%%%%%%%%%%%%%%%%%
\subsection{Classical Monte-Carlo results}
%%%%%%%%%%%%%%%%%%%%%%%%%%%%%%%%%%%%%%%%%%%%%%%%%%%%%%%%%%%%%%%%%%%%%%%

We now assess the validity of the SCGA predictions using the classical Monte-Carlo approach.
For all values of $J_{z\pm} \in (0, \sqrt{2})$ and not too low temperatures, we find both approaches to yield consistent results, as in the non-Kramers case for $J_{z\pm}=0$ discussed above. In particular, the spin structure factor measured in the Monte-Carlo simulations is consistent with the emergence of a classical higher-rank spin liquid regime at intermediate temperatures.
At low temperatures, we observe a finite-temperature transition towards to one out of two possible symmetry-broken states: a $\mathbf q=0$ AIAO antiferromagnet for $J_{z\pm} < 0.5$ and a non-magnetic spin nematic for $J_{z\pm} > 0.5$, see Fig.~\ref{fig:Schematic_phase_diagram_small}.
In the following, we characterize the three states in detail.
%

%%%%%%%%%%%%%%%%%%%%%%%%%%%%%%%%%%%%%%%%%%%%%%%%%%%%%%%%%%%%%%%%%%%%%%%
\subsubsection{Classical spin liquid}
%%%%%%%%%%%%%%%%%%%%%%%%%%%%%%%%%%%%%%%%%%%%%%%%%%%%%%%%%%%%%%%%%%%%%%%

\begin{figure}[tb!]
    \centering
   \begin{overpic}[width=1.\columnwidth]{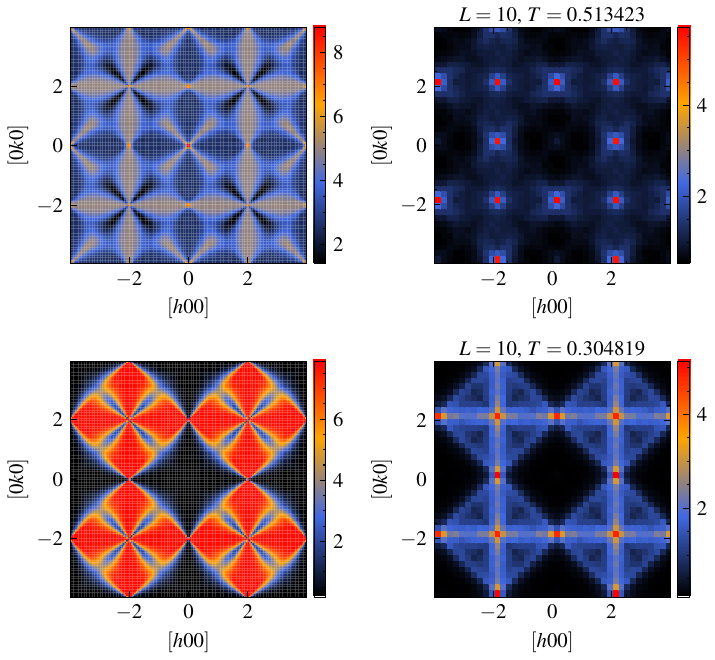}
    \put(5,90){(a)}
    \put(5,44){(c)}
    \put(55,90){(b)}
    \put(55,44){(d)}
    \end{overpic}
    \caption{%
    (a)~Static spin structure factor in the $[hk0]$ plane from SCGA in the low-temperature limit for the Kramers case, using a representative value of $J_{z\pm} = 0.4$, describing the correlations in the rank-2 classical spin liquid.
    (b)~Same as~(a), but from classical Monte-Carlo simulations on an $L=10$ lattice for a fixed temperature $T = 0.51$ above $\Tc$.
    (c)~Same as~(a), but for a value of $J_{z\pm} = 1$.
    (d)~Same as~(c), but from classical Monte-Carlo simulations on an $L=10$ lattice for a fixed temperature $T = 0.30$ above $\Tc$. We note that in all panels fourfold pinch points, or remnants of such correlations, can be observed at the $[220]$ and symmetry related points.
    }
\label{fig:CSL_phase}
\end{figure}

The SCGA analysis predicts a rank-2 U(1) spin liquid phase for $|J_{z\pm}|>0$. Figure~\ref{fig:CSL_phase}(a) shows the corresponding spin structure factor from SCGA for a representative value of $J_{z\pm}=0.4$, indicating the emergence of fourfold (twofold) pinch points at [220] ([020]) and symmetry-related points in the $[hk0]$ plane in reciprocal space. These features can be understood to descend from the Gauss' laws in Eq.~\eqref{eq:gauss_laws_jzpm}.
For intermediate temperatures, our large-scale Monte-Carlo simulations reveal broadened versions of the combined twofold and fourfold pinch-point pattern. The intermediate-temperature regime can thus be understood as a an excited state of the
higher-rank spin liquid found in SCGA, where gauge charges are present and the higher-rank Gauss's law is violated. This is illustrated in Fig.~\ref{fig:CSL_phase}(b), which shows the static spin structure factor for $J_{z\pm}=0.4$ at a fixed finite temperature slightly above the transition temperature.

The rank-2 spin liquid phase at intermediate temperatures predicted by the SCGA analysis can be understood as a single phase for $J_{z\pm}>0$, this however does not imply that all the thermodynamic signatures in this phase remain unchanged as a function of the interaction parameter $J_{z\pm}$. Indeed, the Gauss's laws in Eq.~\eqref{eq:gauss_laws_jzpm} possess an implicit dependence on the $J_{z\pm}$ parameter encoded in the $T_{1-}$ field, see Appendix~\ref{appendix:irreps}. This functional dependence is reflected in the spin structure factors of the corresponding model. Figures~\ref{fig:CSL_phase}(a) and \ref{fig:CSL_phase}(c) show the SCGA predictions for the spin structure factor for $J_{z\pm}=0.4$ and $J_{z\pm}=1$, respectively, both possessing twofold and fourfold pinch points. A similar evolution in the static spin structure factor is observed in the
classical Monte-Carlo simulations, as shown in Figs.~\ref{fig:CSL_phase}(b) and \ref{fig:CSL_phase}(d), for the same set of interaction parameters, but at a fixed finite temperature slightly above the transition temperature.
Indeed, the qualitative agreement between Figs.~\ref{fig:CSL_phase}(b) and \ref{fig:CSL_phase}(d) and their corresponding SCGA predictions indicates that the spin-liquid regime persists above the low-temperature order throughout the phase diagram.
%

%%%%%%%%%%%%%%%%%%%%%%%%%%%%%%%%%%%%%%%%%%%%%%%%%%%%%%%%%%%%%%%%%%%%%%%
\subsubsection{\texorpdfstring{$A_2$}{A2} all-in-all-out antiferromagnet}
%%%%%%%%%%%%%%%%%%%%%%%%%%%%%%%%%%%%%%%%%%%%%%%%%%%%%%%%%%%%%%%%%%%%%%%

\begin{figure}[tb!]
    \centering
   \begin{overpic}[width=1.\columnwidth]{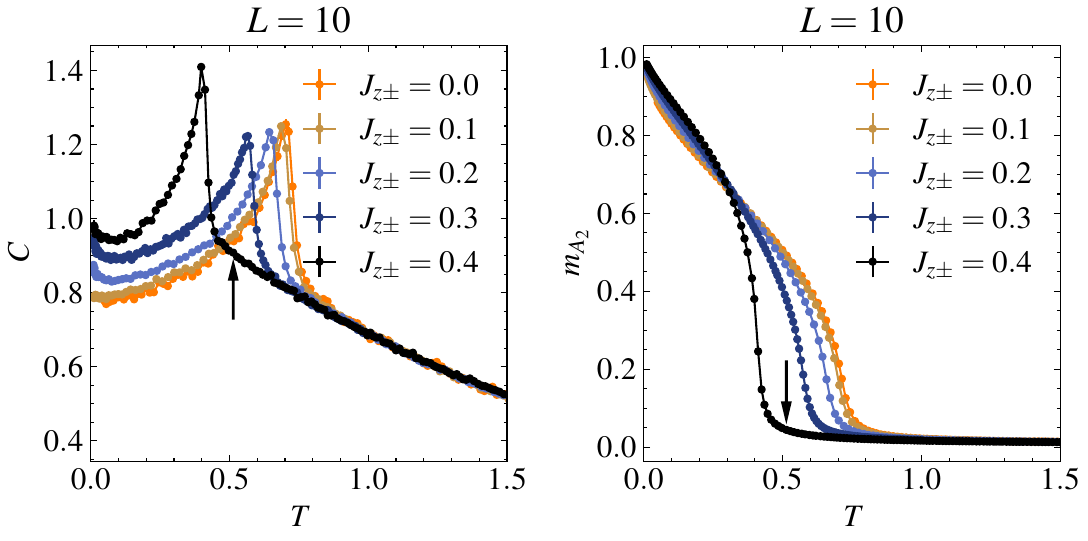}
    \put(5,48){(a)}
    \put(55,48){(b)}
    \end{overpic}
    \caption{%
    (a)~Specific heat as function of temperature $T$ from classical Monte-Carlo simulations on an $L=10$ lattice for different values of $J_{z\pm}$, indicating that $\Tc$ initially decreases with increasing $J_{z\pm}$.
    (b)~Same as (a), but for the $A_2$ order parameter $m_{A_2}$.
    Arrows in (a) and (b) indicate the temperatures chosen for the spin structure factor measured in Fig.~\ref{fig:CSL_phase}(b).
    }
\label{fig:A2_CMC}
\end{figure}

Similarly to the non-Kramers case, we assess the low-temperature fate of the classical spin liquid using the Monte-Carlo approach. For
$J_{z\pm}<0.5$, the low-temperature behavior in the Kramers case resembles that of the non-Kramers case $J_{z\pm}=0$, where a continuous transition into a long-range ordered phase is observed, see Fig.~\ref{fig:A2_CMC}(a), into a $\mathbf q=0$ AIAO antiferromagnetic order characterized the $A_2$ irrep, see Fig.~\ref{fig:A2_CMC}(b).
We note that the precise $\Tc$ depends on $J_{z\pm}$, see Fig.~\ref{fig:A2_CMC}(a).

For finite $J_{z\pm} \in (0, 0.5)$, the behavior of the system within the $A_2$ ordered phase is quite remarkable. As shown in Fig.~\ref{fig:A2_CMC}(a), for decreasing temperature $T$, the specific heat
initially decreases below one and eventually increases again toward one at extremely low temperatures. We note that the upturn in the specific heat below $\Tc$ becomes more prominent for larger $J_{z\pm}$. This behavior can be understood in terms of the classical low-temperature expansion: for $J_{z\pm} > 0$, the four bands that are flat at quadratic order for the $A_2$ state in the non-Kramers case become dispersive, resulting in $n_2 = 8$ and $n_4 = 0$ in Eq.~\eqref{eq:specific-heat-CLTE}, leading to a specific heat of $C = 1$ in the low-temperature limit. For small but finite $J_{z\pm}$, however, the low-energy quadratic modes have a small band width. Thus, for temperatures larger than the band width of the low-energy quadratic modes, the low-energy fluctuations effectively emerge from a flat band, therefore decreasing the value of the specific heat.
On the other hand, as the temperature is further decreased, the weak dispersions of the low-energy bands eventually become relevant, driving the system to a unit specific heat in the low-temperature limit, in agreeance with the equipartition theorem of quadratic modes. As the width of the low-energy dispersive bands increases with $J_{z\pm}$, the characteristic temperature of the upturn should be expected to increase as well, in agreement with the trend observed in Fig.~\ref{fig:A2_CMC}(a), and further shown in Appendix~\ref{appendix:C_v}. A more detailed analysis of the upturn in the specific heat may be obtained by extending the low-temperature expansion beyond the quadratic order. This is left for future work.

%%%%%%%%%%%%%%%%%%%%%%%%%%%%%%%%%%%%%%%%%%%%%%%%%%%%%%%%%%%%%%%%%%%%%%%
\subsubsection{Spin nematic}
%%%%%%%%%%%%%%%%%%%%%%%%%%%%%%%%%%%%%%%%%%%%%%%%%%%%%%%%%%%%%%%%%%%%%%%

\begin{figure}[tb!]
    \centering
    \begin{overpic}[width=1.\columnwidth]{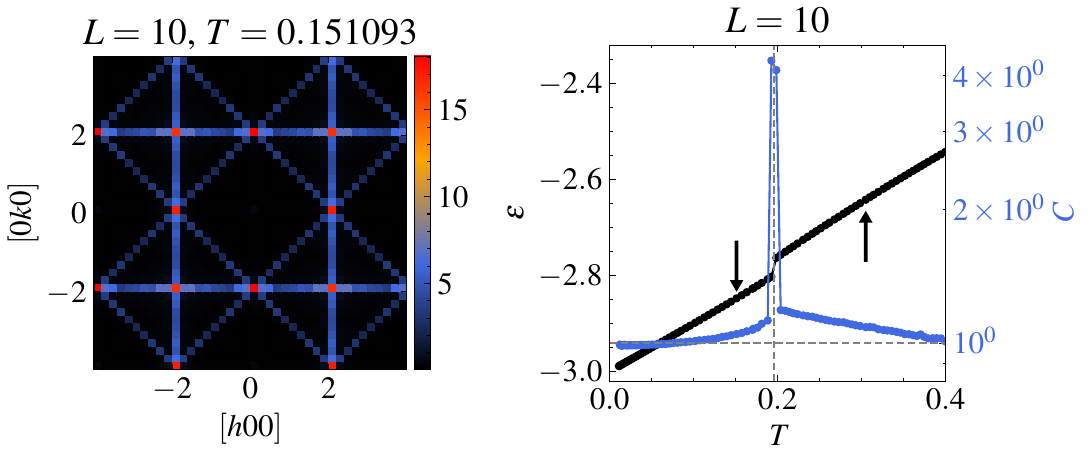}
    \put(1,41){(a)}
    \put(51,41){(b)}
    \end{overpic}
    \caption{%
    (a)~Static spin structure factor in the $[hk0]$ plane from classical Monte-Carlo simulations on an $L=10$ lattice for a representative value of $J_{z\pm} = 1$ and fixed temperature  $T=0.15$ below $\Tc$, describing the correlations in the spin nematic phase.
    (b)~Energy density $\varepsilon$ (black) and specific heat $C$ (blue) as function of temperature for $J_{z\pm} = 1$, revealing the strong first-order nature of the spin-liquid-to-nematic transition, marked by the jump (sharp peak) in $\varepsilon$ ($C$).
    Dashed line indicates the transition temperature. Arrows indicate the temperatures chosen in (a) and Fig.~\ref{fig:CSL_phase}(d).
    }
\label{fig:SCGA_SN_kramers}
\end{figure}

As shown in Fig.~\ref{fig:Schematic_phase_diagram_small}, along the $A_2\oplus T_1\oplus T_2$ line,
an ObD mechanism lifts the classical spin liquid degeneracy at low temperatures, driving the system into a symmetry-breaking low-temperature phase. As shown in the previous subsection, for $J_{z\pm} < 0.5$, the low-temperature symmetry-broken phase is a $\mathbf{q}=0$ AIAO phase.
Remarkably, for $J_{z\pm} > 0.5$, the nature of the low-temperature state changes abruptly.
At low temperatures, the system realizes a novel spin nematic ground state, which lacks any magnetic long-range order.
This is illustrated in Fig.~\ref{fig:SCGA_SN_kramers}(a), which shows the static spin structure factor from classical Monte-Carlo simulations for a representative value of $J_{z\pm} =1$ and a fixed temperature below the transition temperature.
The structure factor is characterized by the absence of any magnetic Bragg peaks and instead shows continuous intensity lines along $[2k0]$ and symmetry-related lines in the $[hk0]$ plane, which persist down to the lowest temperatures. Moreover, we have explicitly verified that all magnetic order parameters $m_I$, with $I \in \{A_2, E, T_2, T_{1\pm}\}$, scale towards zero in the thermodynamic limit in this phase (not shown). These two observations indicate that the low-temperature phase is nonmagnetic with time reversal symmetry remaining intact down to the lowest temperatures.
The symmetry-breaking transition is strongly first order, as revealed by the sharp jump in the energy density shown in Fig.~\ref{fig:SCGA_SN_kramers}(b).
The sharp discontinuity at the transition makes it difficult to numerically extract the specific heat peak as defined in Eq.~\eqref{eq:specific_heat-MC}, i.e., from measuring energy fluctuations. As a more stable numerical approach, for the specific heat data shown in Fig.~\ref{fig:SCGA_SN_kramers}(b), we have used symmetric two-point numerical differentiation of the energy density.
Furthermore, we note that, in the low-temperature limit, the specific heat plateaus towards $1$ for all values of $J_{z\pm} \in (0.5,\sqrt{2})$ along the $A_2\oplus T_1\oplus T_2$ line, suggesting that thermal spin fluctuations about the ground-state configuration are purely quadratic.
We emphasize that the position of the critical temperature $\Tc$ varies with $J_{z\pm}$, as shown in Fig.~\ref{fig:Schematic_phase_diagram_small},
eventually going to zero for $J_{z\pm} \to \sqrt{2}$, corresponding to the termination point of the $A_2\oplus T_1\oplus T_2$ line, identified with the antiferromagnetic Heisenberg model.%
\footnote{This termination point corresponds to a phase boundary between \emph{four} antiferromagnetic orders, namely the $A_2$, $E$, $ T_{1-}$, and $T_2$ orders. The additional degeneracy with the $E$ ordered phase at the termination point results in the transformation of the rank-2 Gauss's laws in Eq.~\eqref{eq:gauss_laws_jzpm} into the three rank-1 Gauss's laws of the antiferromagnetic Heisenberg model~\cite{Moessner1998_low_temp}, ensuring the vanishing of the three components of the total magnetization vector.
}

\begin{figure}[tb!]
    \centering
    \begin{overpic}[width=1\columnwidth]{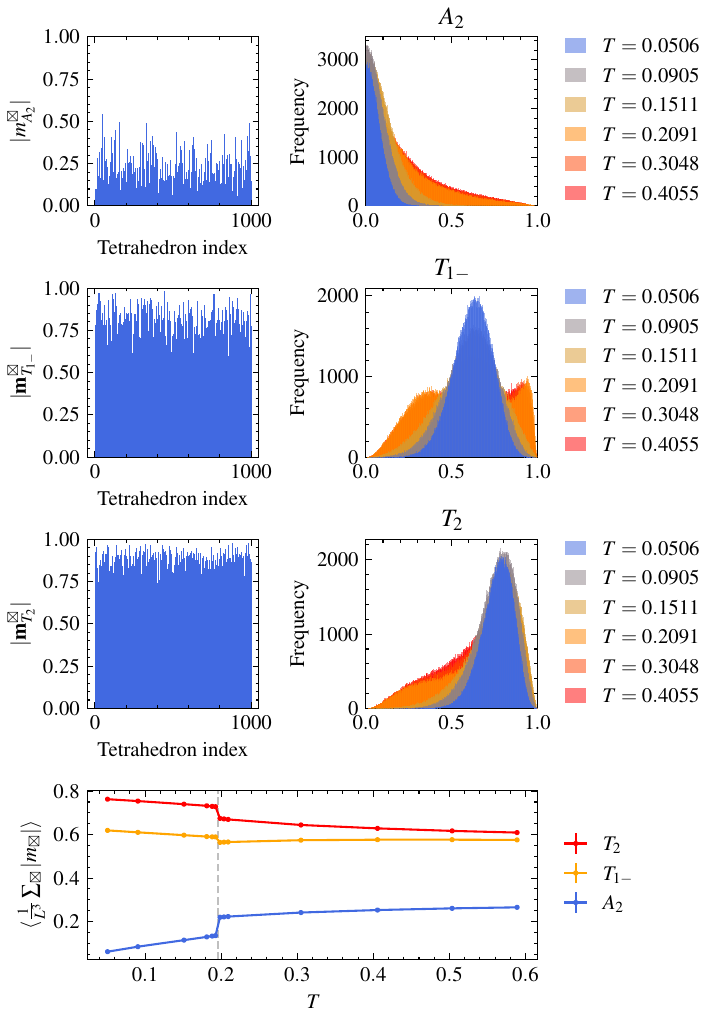}
    \put(0,98){(a)}
    \put(0,73){(b)}
    \put(0,48){(c)}
    \put(0,23){(g)}

    \put(33,98){(d)}
    \put(33,73){(e)}
    \put(33,48){(f)}
    \end{overpic}
    \caption{
    (a)~Distribution of the magnitude of the $A_2$ irrep across the different tetrahedra of an $L=10$ lattice, obtained from a single Monte-Carlo snapshot, randomly selected from a simulation at temperature $T=0.15$ and $J_{z\pm}=1$.
    (b)~Same as~(a), but for the magnitude of the $T_{1-}$ irrep.
    (c)~Same as~(a), but for the magnitude of the $T_{2}$ irrep.
    (d)~Histogram of the magnitude of the $A_2$ irrep across many Monte-Carlo snapshots, obtained for $J_{z\pm} = 1$ and various fixed temperatures above and below $\Tc$, averaged over all tetrahedra for each snapshot.
    (e)~Same as (d), but for the magnitude of the $T_{1-}$ irrep.
    (f)~Same as~(d), but for the magnitude of the $T_{2}$ irrep.
    (g)~Monte-Carlo average of the single-tetrahedron $A_2$, $T_{1-}$, and $T_2$ irreps magnitude as function of temperature. Dashed line indicates the transition temperature.
    }
\label{fig:distributions}
\end{figure}

Since the low-temperature phase for $J_{z\pm} > 0.5$ features no magnetic long-range order, the magnetic order parameters associated with the irreps, as defined in Eq.~\eqref{eq:irreps_measures}, are not suited to characterize this phase. However, the \emph{distributions} of the magnitude of the irreps, calculated at the single-tetrahedron level, represent a useful quantity to understand the nature of the symmetry-broken ground state. Figures~\ref{fig:distributions}(a)-(c) show the $A_2$, $T_{1-}$, and $T_2$ irrep magnitudes on each of the tetrahedra on an $L=10$ lattice for a single Monte-Carlo snapshot, randomly selected from a simulation with $J_{z\pm}=1$ at a temperature $T=0.15$ below the transition temperature $\Tc$.
For this snapshot, the $A_2$ irrep has, on practically all tetrahedra, a significantly smaller magnitude in comparison with the $T_{1-}$ and $T_2$ irreps. This suggests that out of the three competing orders $A_2$, $T_{1-}$, and $T_2$, the low-temperature phase is majorly built from only the $T_{1-}$ and $T_2$ states. This observation can be further corroborated by considering histograms of the irreps across many Monte-Carlo snapshots, obtained from the simulation at a fixed temperature. The histograms of the $A_2$, $T_{1-}$, and $T_2$ irrep magnitudes for $J_{z\pm} = 1$ and different temperatures above and below $\Tc$, averaged over all tetrahedra for each snapshot, are shown in Figs.~\ref{fig:distributions}(d)-(f). Additionally, we report the temperature dependence of the average of these distributions in Fig.~\ref{fig:distributions}(g).
The evolution of the histograms shows that lowering the temperature leads to a thermal depopulation of the $A_2$ irrep, reflected on the corresponding average magnitude of the $A_2$ irrep going to zero, see Figs.~\ref{fig:distributions}(d) and \ref{fig:distributions}(g). In contrast, the $T_{1-}$ and $T_2$ irreps acquire a finite average magnitude as suggested from the well-defined peaks in the histograms centered at non-zero values.
The behavior of the distributions suggests using the average of the single-tetrahedron irreps magnitude as an effective quantity to signal the onset of the low-temperature phase. Indeed, Fig.~\ref{fig:distributions}(g) reveals that these quantities indeed feature a clear jump around the transition temperature, with decreasing $A_2$, and increasing $T_2$ and $T_{1-}$.%
\footnote{It is crucial to consider the magnitude $|\mathbf{m}_I^\boxtimes|$ of the irrep at the level of the single tetrahedron before averaging over the lattice, since $|\sum_\boxtimes \mathbf{m}_I^\boxtimes|$ vanishes in the thermodynamic limit due to the unbroken time reversal symmetry of the low-temperature state.}

\begin{figure}[tb!]
    \centering
    \begin{overpic}[width=1\columnwidth]{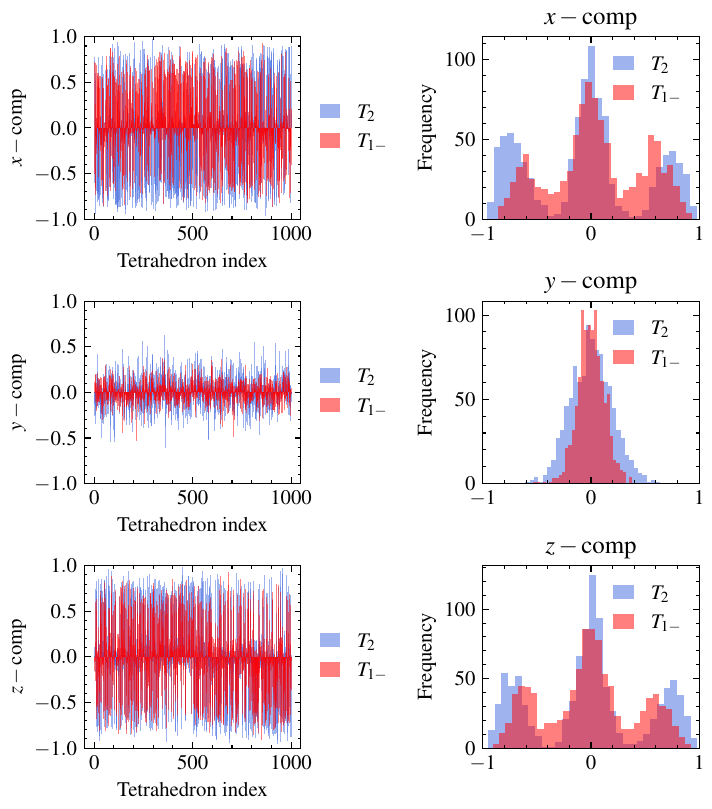}
    \put(2,97.5){(a)}
    \put(2,64){(b)}
    \put(2,31){(c)}

    \put(54,97.5){(d)}
    \put(54,64){(e)}
    \put(54,31){(f)}
    \end{overpic}
    \caption{%
    (a)~Distribution of the $x$ component of the $T_{1-}$ and $T_2$ irrep across the different tetrahedra of an $L=10$ lattice, obtained from a single Monte-Carlo snapshot, randomly selected from a simulation at $T=0.15$ and $J_{z\pm} = 1$.
    (b)~Same as~(a), but for the $y$ components.
    (c)~Same as~(a), but for the $z$ components.
    (d)-(f)~Histograms of distributions shown in panels~(a)-(c).
    }
\label{fig:distributions_components}
\end{figure}

The low-temperature state composed of $T_2$ and $T_{1-}$ irreps is characterized by a specific behavior of the irreps components $T_{1-}^\alpha$ and $T_{2}^\alpha$ with $\alpha\in\{x,y,z\}$.
Figures~\ref{fig:distributions_components}(a)-(c) show the distributions of the three different components of $T_2$ and $T_{1-}$ irreps across the different tetrahedra of an $L = 10$ lattice, obtained from a single Monte-Carlo snapshot, randomly selected from a simulation at temperature $T = 0.15$ and $J_{z\pm} = 1$.
The corresponding histograms across the different tetrahedra for the same Monte-Carlo snapshot are shown in Figs.~\ref{fig:distributions_components}(d)-(f).
As is evident from these figures, one out of the three components $\{x,y,z\}$ of each of the two irreps is suppressed,
where the suppressed component is the same for
both irreps.
For the Monte-Carlo snapshot shown, the component $y$ has a smaller magnitude on each tetrahedron, and the corresponding histogram [Fig.~\ref{fig:distributions_components}(e)] is peaked around zero. The asymmetry between the distributions of the three irrep components indicates a spontaneous breaking of the cubic symmetry in the low-temperature phase.
We have explicitly verified that other Monte-Carlo snapshots feature an analogous behavior, where the suppressed component varies between different snapshots, as required from the cubic symmetry of the model. The ground state manifold therefore falls into three disjoint sets characterized by suppressed $x$, $y$, or $z$ components of the $T_2$ and $T_{1-}$ irreps.
However, each of these three sets contains a macroscopic number of degenerate states, reflected in the
static spin structure factor, characterized by the absence of Bragg peaks and the presence of continuous intensity lines.
For each set, these lines are oriented along a single direction, reflecting the spontaneous breaking of cubic spin symmetry.%
\footnote{The three sets of lines in Fig.~\ref{fig:SCGA_SN_kramers}(a) arise in our classical Monte-Carlo simulations due to the parallel tempering algorithm, which satisfies detailed balance and therefore uniformly samples across all symmetry-broken states.}
The extensive degeneracy in this phase can be exposed by studying the ground-state manifold whose states have all tetrahedra in a spin configuration defined by the $T_{1-}$ and $T_2$ irreps.
In the zero-temperature limit, for each tetrahedron, a continuous ground-state manifold can be constructed by linearly combining the \emph{same} components of both the $T_{1-}$ and $T_2$ irreps, resulting in an accidental U(1) symmetry. The spin modes in this manifold are parametrized by $\mathrm{m}^\alpha_{\perp}(\theta) = \cos\theta\ \mathrm{m}^\alpha_{T_{1-}} + \sin\theta\ \mathrm{m}^\alpha_{T_{2}}$ for $\theta \in [0, 2\pi)$, which is analogous to the construction described in Ref.~\cite{Noculak_HDM_2023} involving a component of the $E$ and $T_{1-}$ irreps.
To construct a lattice configuration from the $\mathrm{m}_\perp^\alpha$ single-tetrahedron states, one must tile these states while respecting the fact that adjacent tetrahedra share one spin.
A straightforward lattice configuration resulting from this approach is a $\textbf{q}=0$ configuration, where all ``up'' tetrahedra adopt the same spin configuration $\mathrm{m}_\perp^\alpha(\theta)$. Alternatively, a \emph{mixed} spin configuration can be created by tiling two distinct $\mathrm{m}_\perp(\theta)$ spin configurations, such as $\mathrm{m}_\perp^\alpha(\theta_\alpha)$ and $\mathrm{m}_\perp^\beta(\theta_\beta)$, where $\alpha\neq\beta$.%
\footnote{Constructing a mixed state with three spatial components is not feasible because the $\theta$ angles required to tile two distinct $\mathrm{m}_\perp^\alpha$ states are not compatible, resulting in higher-energy configuration.}
This tiling requires specific values for the angles $\theta_\alpha$ and $\theta_\beta$, as evidenced by the intermediate values of the irrep components in the distributions shown in Figs.~\ref{fig:distributions_components}(d) and \ref{fig:distributions_components}(f). We leave further details of the construction of full lattice configurations for future work.

\begin{figure}[tb!]
    \centering
    \begin{overpic}[width=0.6\columnwidth]{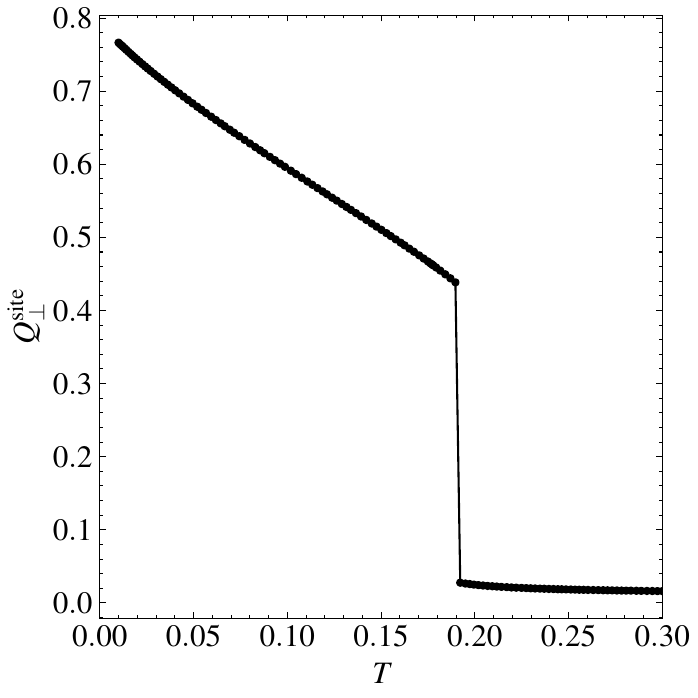}
    \end{overpic}
    \caption{Single-site nematic order parameter $\mathbf{Q}_{\perp}^{\mathrm{site}}$ as function of temperature for $J_{z\pm = 1}$ from classical Monte-Carlo simulations on an $L=10$ lattice. The abrupt change at $\Tc = 0.191(1)$ signals the onset of spin nematic order with a local $xy$ nature.}
\label{fig:nematic_op}
\end{figure}

The above analysis suggests that every single-tetrahedron configuration in the low-temperature phase selects a specific angle $\theta$ of the accidental U(1) symmetry ground-state manifold generated by the degeneracy of the $T_{1-}$ and $T_2$ irreps. This selection lifts the accidental U(1) symmetry, which is also reflected on a cubic symmetry breaking associated to the thermal depopulation of one of the components of the $T_{1-}$ and $T_2$ irreps, see Fig.~\ref{fig:distributions_components}. The synchronous breaking of the cubic symmetry and the accidental U(1) symmetry suggests that the low-temperature phase can be understood as a spin nematic with a local $xy$ nature. In order to demonstrate that this is indeed the case, we show in Fig.~\ref{fig:nematic_op} the single-site spin nematic order parameter $\mathbf{Q}_{\perp}^{\mathrm{site}}$, defined in Eq.~\eqref{eq:single_site_nematic_op}, as function of temperature $T$, for a fixed representative value of $J_{z\pm} = 1$.
The nematic order parameter
vanishes in the paramagnetic phase and exhibits a distinct jump to a finite value at the first-order transition.
Consequently, the low-temperature phase forms a spin nematic characterized by the breaking of cubic rotational symmetry. In this phase, spins fluctuate within the $T_2$ and $T_{1-}$ irreps on local $xy$ planes, while time-reversal symmetry remains intact down to the lowest temperatures.

%%%%%%%%%%%%%%%%%%%%%%%%%%%%%%%%%%%%%%%%%%%%%%%%%%%%%%%%%%%%%%%%%%%%%%%
\subsection{Exposing the spin nematic using SCGA}
%%%%%%%%%%%%%%%%%%%%%%%%%%%%%%%%%%%%%%%%%%%%%%%%%%%%%%%%%%%%%%%%%%%%%%%

\begin{figure}[tb!]
    \centering
    \begin{overpic}[width=1.\columnwidth]{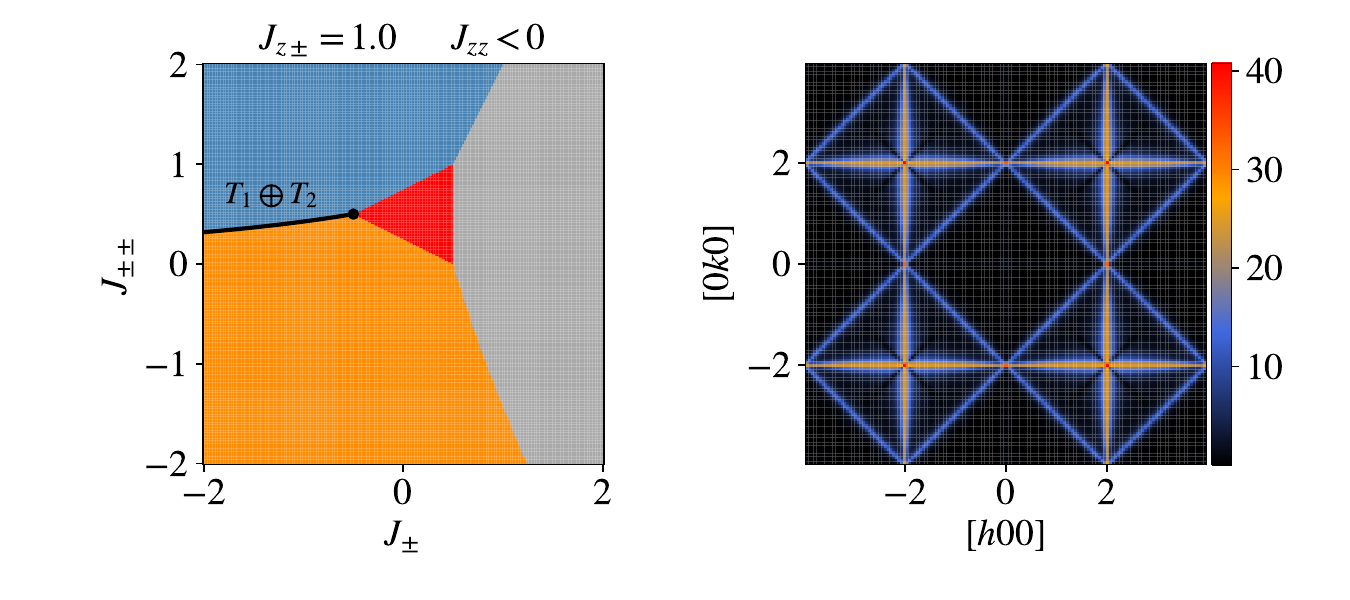}
    \put(6,43){(a)}
    \put(53,43){(b)}
    % 44,57.5
    \put(18,23.5){\fontsize{5}{38} $A_2\oplus T_{1}\oplus T_2$}
    \put(24.8,26.5){\fontsize{5}{38} $\uparrow$}
    \put(21,15){\fontsize{7}{38} $T_{1-}$}
    \put(38,28){\fontsize{7}{38} $E$}
    \put(28,28){\fontsize{7}{38} $A_2$}
    \put(21,35){\fontsize{7}{38} $T_2$}
    \end{overpic}
    \caption{%
    (a)~Ground state phase diagram of the full model in Eq.~\eqref{eq:general_bilinear_hamiltonian} as function of $J_{\pm}$ and $J_{\pm\pm}$ for fixed $J_{z\pm}=1$ and $J_{zz}=-1$.
    The $A_2 \oplus T_1 \oplus T_2$ line pierces the $J_{\pm}$-$J_{\pm\pm}$ phase diagram at the left-most corner of the $A_2$ phase, as indicated by the black dot.
    The intersection of the $T_{1}\oplus T_2$ plane with the $J_{\pm}$-$J_{\pm\pm}$ phase diagram is marked as black line.
    (b)~Static spin structure factor in the $[hk0]$ plane from SCGA in the low-temperature limit for the point $J_{z\pm} = 1$ and $J_{\pm} =-0.625$ on the $T_{1}\oplus T_2$ plane, i.e., for $J_{\pm\pm}$ chosen according to Eq.~\eqref{eq:parametrization_T1_T2}, describing the correlations in the spin nematic phase.
    }
\label{fig:SCGA_nematic}
\end{figure}

The spin nematic phase observed in our classical Monte-Carlo simulations is not captured by a plain SCGA analysis at the $A_2 \oplus T_1 \oplus T_2$ line. In fact, in the plain SCGA analysis, the $A_2$ irrep is predicted to be thermally populated down to the lowest temperatures, in stark contrast to its sudden depopulation observed in the Monte-Carlo simulations. The depopulation of the $A_2$ irrep observed in the classical Monte-Carlo simulations suggests that the Ginzburg-Landau free energy functional at the $A_2 \oplus T_1 \oplus T_2$ line for $J_{z\pm}>0.5$ contains higher-order terms, not captured by the SCGA analysis, that suppress the $A_2$ irrep for $J_{z\pm}>0.5$.
A naive way of including the depopulation of the $A_2$ irrep in the SCGA analysis would be to introduce a small energy cost to the $A_2$ irrep. However, simply adding an additional gap would neither reproduce the spin structure factors observed in the Monte-Carlo simulations nor expose the underlying physics of the selection.
As an alternative approach, we investigate spin models defined in the vicinity of the $A_2 \oplus T_1 \oplus T_2$ line
in a region where the $A_2$ irrep becomes gapped, while the $T_{1-}$ and $T_2$ irreps remain degenerate.
To this end, we relax the restrictions on the couplings $J_{\pm}$ and $J_{\pm\pm}$ in Eq.~\eqref{eq:parametrization_A2_T1_T2} in order to parametrize a plane in coupling space at which the $T_{1-}$ and $T_{2}$ states compete.
Figure~\ref{fig:SCGA_nematic}(a) shows the ground state phase diagram of the full model in Eq.~\eqref{eq:general_bilinear_hamiltonian} as function of $J_{\pm}$ and $J_{\pm\pm}$ for fixed $J_{z\pm}=1$ and $J_{zz}=-1$. The $A_2 \oplus T_1 \oplus T_2$ line discussed so far pierces the $J_{\pm}$-$J_{\pm\pm}$ phase diagram for fixed $J_{z\pm}=1$ at the left-most corner of the $A_2$ phase, as indicated by the black dot in Fig.~\ref{fig:SCGA_nematic}(a).
The $A_2 \oplus T_1 \oplus T_2$ line connects to a $T_1 \oplus T_2$ plane%
\footnote{Due to the constraints in Eqs.~\eqref{eq:constraints_A2_T1_T2} and \eqref{eq:constraints_T1_T2}, the $T_1 \oplus T_2$ ``plane'' is, geometrically, rather an infinitely-long finite-width strip with two parallel line boundaries in the width direction [Eq.~\eqref{eq:constraints_A2_T1_T2}] and a single parabolic boundary in the length direction [Eq.~\eqref{eq:constraints_T1_T2}].}
located at the boundary between the $T_{1-}$ and $T_2$ phases. The intersection between the $T_{1}\oplus T_2$ plane and the $J_{\pm}$-$J_{\pm\pm}$ phase diagram is indicated by the black line in Fig.~\ref{fig:SCGA_nematic}(a).
The $T_1 \oplus T_2$ plane can be parametrized as function of $J_{\pm}$ and $J_{z\pm}$ by
\begin{eqnarray} \label{eq:parametrization_T1_T2}
    J_{\pm\pm}=\frac{1}{8}\left(2J_{\pm}+J_{zz}+\sqrt{4J_{\pm}^2+32J_{z\pm}^2+4J_{\pm}J_{zz}+4J_{zz}^2}\right),\nonumber\\
\end{eqnarray}
where
\begin{eqnarray} \label{eq:constraints_T1_T2}
    J_{\pm}<-\frac{J_{z\pm}^2}{J_{zz}}+\frac{3}{2}J_{zz}.
\end{eqnarray}

Figure~\ref{fig:SCGA_nematic}(b) shows the static spin structure factor for a representative point on the $T_{1}\oplus T_2$ plane obtained via SCGA.
Remarkably, it exhibits continuous intensity lines just as those seen by the classical Monte-Carlo simulations in the spin nematic phase, cf.\ Fig.~\ref{fig:SCGA_SN_kramers}(a).
We have explicitly verified that these features of the spin structure factor remain largely unchanged throughout the $T_{1}\oplus T_2$ plane.\footnote{This excludes the case for which $J_{z\pm}=0$ discussed in Ref.~\cite{Taillefumier2017_xxz}.}
This suggests that the spin nematic phase observed at low temperatures in the classical Monte-Carlo simulations at the $A_2 \oplus T_{1}\oplus T_2$ line for $J_{z\pm} > 0.5$ extends over the whole $T_{1}\oplus T_2$ plane.
In fact, we have explicitly confirmed this prediction of the SCGA analysis by classical Monte-Carlo simulations at a representative point $\{J_{zz},J_{\pm},J_{\pm\pm},J_{z\pm}\}=\{-1,-0.7,\frac{\sqrt{59}-3}{10},1\}$ on the $T_{1}\oplus T_2$ plane. The identification of the extended stability region of the spin nematic phase enables us to reinterpret the physics along the $A_2 \oplus T_1 \oplus T_2$ line as a competition between the $A_2$ antiferromagnet, stabilized within the central red triangle in Fig.~\ref{fig:SCGA_nematic}(a), and the spin nematic, stabilized along the $T_1 \oplus T_2$ plane. The outcome of this competition yields a classical spin liquid at intermediate temperatures, which ultimately is destabilized by a thermal ObD selection of a symmetry-breaking phase at low temperatures. The phase selected depends on the value of $J_{z\pm}$: our classical Monte-Carlo simulations demonstrate that the $A_2$ phase is thermally selected for $J_{z\pm} < 0.5$, while the spin nematic phase is favored for $J_{z\pm} > 0.5$.

Having established the relation between the low-temperature phase observed along the $A_2 \oplus T_{1}\oplus T_2$ line and the low-temperature phase in the $T_{1}\oplus T_2$ plane, we use the SCGA analysis of the $T_{1}\oplus T_2$ plane to rationalize the origin of the continuous intensity lines observed in the static spin structure factors in Figs.~\ref{fig:SCGA_SN_kramers}(b) and \ref{fig:SCGA_nematic}(b). Away from the $A_2 \oplus T_{1}\oplus T_2$ line within the $T_{1}\oplus T_2$ plane, the low-energy bands in the spectrum of the interaction matrix shown in Fig.~\ref{fig:bands} acquire a finite dispersion in the $[hk0]$ plane for all non-zero values of $J_{z\pm}$, but remain flat along the $[hhh]$, $[h00]$, and symmetry-related directions in reciprocal space.%
\footnote{Interestingly, the bands of this model remain flat in the $[hh\ell]$ plane.}
The existence of such flat lines in the spectrum of the interaction matrix implies that the degeneracy of the ground-state manifold for this phase scales at least linearly in size. We emphasize that the low-energy flat lines in the band spectrum are observed for \emph{all} non-zero values of $J_{z\pm}$ and $J_{\pm}$ in the $T_{1}\oplus T_2$ plane. The low-energy flat lines in the spectra result in continuous lines of scattering in the SCGA spin structure factors, previously referred to as pinch lines~\cite{Benton2016_Pinch_line}, and observed in Fig.~\ref{fig:SCGA_nematic}(b).

%

%%%%%%%%%%%%%%%%%%%%%%%%%%%%%%%%%%%%%%%%%%%%%%%%%%%%%%%%%%%%%%%%%%%%%%%
\section{Conclusions}
\label{sec:discussion_conclusion}
%%%%%%%%%%%%%%%%%%%%%%%%%%%%%%%%%%%%%%%%%%%%%%%%%%%%%%%%%%%%%%%%%%%%%%%

In this work, we have studied the phase boundary between three $\mathbf{q}=0$ states in the parameter space of a general bilinear nearest-neighbor Hamiltonian on the pyrochlore lattice. Such triple phase boundaries have recently attracted considerable attention as potential hosts for a variety of classical spin liquid phases~\cite{Taillefumier2017_xxz,lozano-gomez2023,atlas_2024}. Applying different analytical and numerical tools, we have constructed the finite-temperature phase diagram along the boundary between the $A_2$ (AIAO antiferromagnet), $T_{1-}$ (splayed ferromagnet), and $T_2$ (Palmer-Chalker antiferromagnet) phases. For a fixed value of $J_{zz}$, this $A_2 \oplus T_1 \oplus T_2$ line can be parametrized by the interaction parameter $J_{z\pm}$. At high temperatures, and for all the models on the $A_2 \oplus T_1 \oplus T_2$ line, we observe a classical spin liquid regime at intermediate temperatures, where pinch-point features in the static spin structure factor develop. Through a self-consistent Gaussian approximation and an irreducible-representation analysis, we show that these spin liquid regimes are described by a rank-1 U(1) theory for $J_{z\pm}=0$, and a rank-2 U(1) theory for $J_{z\pm}\neq 0$. For all the models on the $A_2 \oplus T_1 \oplus T_2$ line, the spin liquid phase is destabilized at low temperatures by an order-by-disorder mechanism, resulting in the selection of a symmetry-broken phase. Using large-scale classical Monte-Carlo simulations, we have characterized the symmetry-broken phases at low temperatures in detail and computed the finite-temperature phase diagram along the entire $A_2 \oplus T_1 \oplus T_2$ line. We have identified two distinct low-temperature phases, one for $J_{z\pm} < 0.5$ and another one for $J_{z\pm} > 0.5$. For $J_{z\pm}< 0.5$, we observe a continuous transition at a critical temperature $\Tc$ towards an AIAO $\mathbf{q}=0$ antiferromagnet. The stabilization of a single $\mathbf{q}=0$ phase below $\Tc$ down to the lowest temperatures is a scenario similar to those observed for  $\rm Er_2 Ti_2 O_7$~\cite{zhitomirsky2012,savary2012} and $\rm Yb_2 Ge_2 O_7$~\cite{Sarkis2020YbGeO}. For $J_{z\pm}\geq0.5$ we identify a novel spin nematic phase at low temperatures, characterized by continuous lines of intensity in the spin structure factor. In contrast to the low-$J_{z\pm}$ case, the transition towards the spin nematic is strongly first order.
The nematic phase features fluctuating components of the $T_{1-}$ and $T_2$ irreps, with all other modes being thermally depopulated. It exhibits three macroscopically-degenerate sets of ground states mutually associated by cubic symmetry. The three distinct sets are characterized by the suppression of a different component of the $T_{1-}$ and $T_2$ irreps in each case.

The identification of the order-by-disorder-selected nematic phase along the $A_2 \oplus T_1 \oplus T_2$ line led us to uncover a two-dimensional $T_{1-} \oplus T_2$ plane in coupling space, spanned by the parameters $J_{z\pm}$ and $J_{\pm}$, along which the nematic state is stabilized.
To the best of our knowledge, the nematic phase discovered along this line represents the first instance of a non-magnetic, yet symmetry-broken state arising from a low-temperature instability of a higher-rank classical spin liquid state.
In the limit $J_{z\pm} \to 0$, it connects to the nematic state observed as the low-temperature instability of the rank-1 classical spin liquid in Ref.~\cite{Taillefumier2017_xxz}.
Our work thus significantly expands the range of magnetic phases resultant of an entropic selection, which otherwise previously only included $\mathbf{q}=0$ phases~\cite{chern10,savary2012,Noculak_HDM_2023}, and in the most exotic cases certain classical spin liquids~\cite{lozano-gomez2023}.

In the future, it would be interesting to study the effects of quantum fluctuations on the nematic state found in the classical limit. This question could be approached by pseudo-fermion or pseudo-majorana functional renormalization group approaches~\cite{BenediktPhysRevB.109.195109,Nils2023PRL,Muller_2024,NoculakPRB2024}, or potentially within a high-temperature series expansion~\cite{oitmaa_book,Oitmaa-2013}. Quantum effects could introduce an even more increased level of complexity in the competition between the various states involved, potentially leading to further exotic phases of matter.

Our results for the $A_2 \oplus T_1 \oplus T_2$ line suggest that similarly rich physics may be at work also at other triple phase boundaries, such as the $A_2 \oplus T_2 \oplus T_E$ line, corresponding to the upper corner of the red triangle in Fig.~\ref{fig:SCGA_nematic}(a), or the $A_2 \oplus T_1 \oplus T_E$ line, corresponding to the lower right corner of the red triangle in Fig.~\ref{fig:SCGA_nematic}(a). These questions will be addressed in a separate paper.
%
%

% \newpage
%%%%%%%%%%%%%%%%%%%%%%%%%%%%%%%%%%%%%%%%%%%%%%%%%%%%%%%%%%%%%%%%%%%%%%%
\appendix
%%%%%%%%%%%%%%%%%%%%%%%%%%%%%%%%%%%%%%%%%%%%%%%%%%%%%%%%%%%%%%%%%%%%%%%

%%%%%%%%%%%%%%%%%%%%%%%%%%%%%%%%%%%%%%%%%%%%%%%%%%%%%%%%%%%%%%%%%%%%%%%
\section{Irreducible representations}
\label{appendix:irreps}
%%%%%%%%%%%%%%%%%%%%%%%%%%%%%%%%%%%%%%%%%%%%%%%%%%%%%%%%%%%%%%%%%%%%%%%

In this appendix, we provide the explicit form of the irreducible representation spin modes and their single-tetrahedron energies as a function of the interaction couplings $\{J_{zz},J_{\pm},J_{\pm\pm},J_{z\pm}\}$.
The five single-tetrahedron irreps $\mathbf{m}_I$ are given in terms of spins by
 \begin{align}
     \mathrm{m}_{A_2} &= \frac{1}{4} \sum_{\mu} \mathbf{S}_{\mu} \cdot \hat{\mathbf{z}}_{\mu}, \\
     \mathbf{m}_{E} &= \frac{1}{4} \sum_{\mu} \begin{pmatrix}
         \mathbf{S}_{\mu} \cdot \hat{\mathbf{x}}_{\mu} \\
         \mathbf{S}_{\mu} \cdot \hat{\mathbf{y}}_{\mu} \\
     \end{pmatrix}, \\
     \mathbf{m}_{T_2} &= \frac{1}{4}\sqrt{\frac{3}{2}} \sum_{\mu} \hat{\mathbf{z}}_{\mu} \times \mathbf{S}_{\mu}, \\
     \mathbf{m}_{T_1,\parallel} &= \frac{1}{4} \sum_{\mu} \mathbf{S}_{\mu}, \\
     \mathbf{m}_{T_1,\perp} &= \frac{\sqrt{3}}{4} \sum_{\mu} \begin{pmatrix}
         \hat{z}_{\mu}^{x} \mathbf{v}_{\mu}^{yz}\cdot \mathbf{S}_{\mu} \\\\
         \hat{z}_{\mu}^{y} \mathbf{v}_{\mu}^{xz}\cdot \mathbf{S}_{\mu} \\\\
         \hat{z}_{\mu}^{z} \mathbf{v}_{\mu}^{xy}\cdot \mathbf{S}_{\mu} \\
     \end{pmatrix},
 \end{align}
where $\mu=0,\dots,3$ specifies the sublattice index for each point in the tetrahedron~\cite{Yan2017_theory_multiphase}. The normalization is chosen to give a unit magnitude if the corresponding irrep is realized. The vectors $\hat{\mathbf{x}}_{\mu},\hat{\mathbf{y}}_{\mu},\hat{\mathbf{z}}_{\mu}$ refer to the local orthonormal cubic coordinate frame associated to the site of sublattice $\mu$. For example, for $\mu=0$ we have
\begin{equation}
    \label{eq:local_coordinate_frame_0_sublattice}
    \hat{\mathbf{x}}_{0} = \frac{1}{\sqrt{6}}\begin{pmatrix}
        -1 \\ -1 \\ 2 \\
    \end{pmatrix},
    \quad
    \hat{\mathbf{y}}_{0} = \frac{1}{\sqrt{2}}\begin{pmatrix}
        1 \\ -1 \\ 0 \\
    \end{pmatrix},
    \quad
    \hat{\mathbf{z}}_{0} = \frac{1}{\sqrt{3}}\begin{pmatrix}
        1 \\ 1 \\ 1 \\
    \end{pmatrix},
\end{equation}
where the components are expressed in a global cartesian coordinate frame. The vectors $\mathbf{v}^{\alpha\beta}$ are the normalized bond vectors attached to a sublattice $\mu$ and lying on the $\alpha$-$\beta$ plane,
\begin{align}
\begin{aligned}
    \mathbf{v}_0^{xy} &=\frac{1}{\sqrt{2}} \begin{pmatrix}
        1 \\ 1 \\ 0 \\
    \end{pmatrix},
    &
    \mathbf{v}_1^{xy} & = - \mathbf{v}_0^{xy}, \\
    \mathbf{v}_2^{xy} &=\frac{1}{\sqrt{2}} \begin{pmatrix}
        1 \\ -1 \\ 0 \\
    \end{pmatrix},
    &
    \mathbf{v}_3^{xy} & = - \mathbf{v}_2^{xy}, \\
    \mathbf{v}_0^{xz} &=\frac{1}{\sqrt{2}} \begin{pmatrix}
        1 \\ 0 \\ 1 \\
    \end{pmatrix},
    &
    \mathbf{v}_2^{xz} & = - \mathbf{v}_0^{xz}, \\
    \mathbf{v}_1^{xz} &=\frac{1}{\sqrt{2}} \begin{pmatrix}
        1 \\ 0 \\ -1 \\
    \end{pmatrix},
    &
    \mathbf{v}_3^{xz} & = - \mathbf{v}_1^{xz}, \\
    \mathbf{v}_0^{yz} &=\frac{1}{\sqrt{2}} \begin{pmatrix}
        0 \\ 1 \\ 1 \\
    \end{pmatrix},
    &
    \mathbf{v}_3^{yz} & = - \mathbf{v}_0^{yz}, \\
    \mathbf{v}_1^{yz} &=\frac{1}{\sqrt{2}} \begin{pmatrix}
        0 \\ 1 \\ -1 \\
    \end{pmatrix},
    &
    \mathbf{v}_2^{yz} & = - \mathbf{v}_1^{yz}.
\end{aligned}
\end{align}
The generic single-tetrahedron Hamiltonian can be written as~\cite{Yan2017_theory_multiphase}
\begin{align}
\label{eq:parallel-perp-hamiltonian}
    \mathcal{H}_\boxtimes   &=\frac{1}{2}\left[
    a_{A_2} \mathrm{m}_{A_2}^2 + a_{E} \mathbf{m}_{E}^2 + a_{T_2} \mathbf{m}_{T_2}^2\right.
    \nonumber\\ & \left. \quad
     + a_{T_{1,\parallel}} \mathbf{m}_{T_{1,\parallel}}^2
    + a_{T_{1,\perp}} \mathbf{m}_{ T_{1,\perp}}^2
    + a_{T_{1,\rm mix}} \mathbf{m}_{T_{1,\parallel}} \cdot \mathbf{m}_{T_{1,\perp}} \right],
\end{align}
where each irreducible representation is associated with an energy $a_I$. Note that the coefficient $a_{T_{1,\rm mix}}$ quantifies the mixing between the two $T_1$ irreps. Each coefficient in Eq.~\eqref{eq:parallel-perp-hamiltonian} is expressed as a function of the couplings as~\cite{Benton2014_thesis}
\begin{equation}
    \begin{aligned}
        a_{A_2} &= 3J_{zz}, \quad a_{E} = -6J_{\pm}, \quad a_{T_2} = 2J_{\pm}-4J_{\pm\pm}, \\
        a_{T_1,\parallel} &= \frac{1}{3} \left(-J_{zz} + 4J_{\pm} +8J_{\pm\pm} + 8\sqrt{2}J_{z\pm} \right), \\
        a_{T_1,\perp} &= \frac{1}{3} \left(-2J_{zz} + 2J_{\pm} +4J_{\pm\pm} - 8\sqrt{2}J_{z\pm} \right), \\
        a_{T_{1,\rm mix}} &= -\frac{\sqrt{2}}{3} \left(-J_{zz} - 2J_{\pm} -4J_{\pm\pm} +2\sqrt{2}J_{z\pm} \right).
    \end{aligned}
\end{equation}
In order to avoid the mixing between the two $T_1$ irreps, one can define a new rotated irrep basis, namely
\begin{equation}
    \begin{pmatrix}
        \mathbf{m}_{T_{1-}} \\ \mathbf{m}_{T_{1+}}
    \end{pmatrix}
    =
    \begin{pmatrix}
        \cos{\phi} & -\sin{\phi} \\
        \sin{\phi} & \cos{\phi} \\
    \end{pmatrix}
    \begin{pmatrix}
        \mathbf{m}_{T_{1},\parallel} \\ \mathbf{m}_{T_{1},\perp}
    \end{pmatrix},
\end{equation}
where the rotation angle is defined by
\begin{equation}
    \label{eq:rotation-angle}
    \tan{2\phi} =\frac{a_{T_{1,\rm mix}}}{a_{T_1,\perp}-a_{T_1,\parallel}}.
\end{equation}
Following the transformation, the Hamiltonian takes the form shown in Eq.~\eqref{eq:irrep_decomp}, with no mixing terms in the $T_1$ irreps and single-tetrahedron energies given by
\begin{equation}
    a_{T_{1\pm}} = \frac{1}{2}\left( a_{T_1,\perp}+a_{T_1,\parallel} \pm \sqrt{(a_{T_1,\perp}-a_{T_1,\parallel})^2 + a_{T_{1,\rm mix}}^2} \right).
\end{equation}

%%%%%%%%%%%%%%%%%%%%%%%%%%%%%%%%%%%%%%%%%%%%%%%%%%%%%%%%%%%%%%%%%%%%%%%
\section{Classical low-temperature expansion}
\label{appendix:CLTE}
%%%%%%%%%%%%%%%%%%%%%%%%%%%%%%%%%%%%%%%%%%%%%%%%%%%%%%%%%%%%%%%%%%%%%%%

\begin{figure}[tb!]
    \centering
    \begin{overpic}[width=0.8\columnwidth]{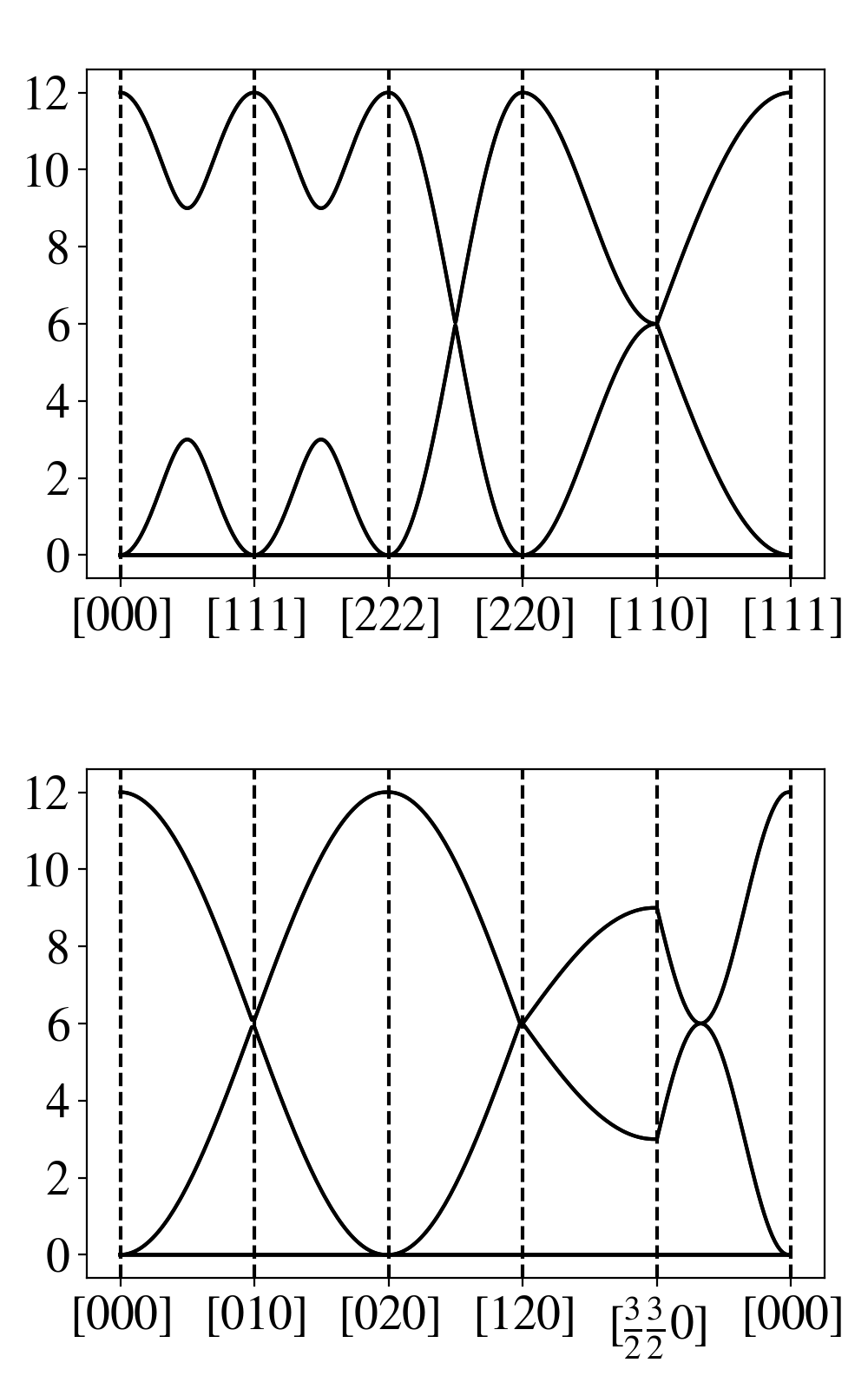}
    \put(5,97){(a)}
    \put(5,47){(b)}
    \put(-4,75){\rotatebox{90}{$E^{(2)}(\mathbf{q} )$}}
    \put(-4,25){\rotatebox{90}{$E^{(2)}(\mathbf{q})$}}
    \end{overpic}
    \caption{Spectrum of the effective Hamiltonian $\mathcal H^{(2)}$ in the classical low-temperature expansion about the $A_2$ phase in the non-Kramers case for $J_{z\pm} = 0$
    along high-symmetry paths in
    (a)~the $[hh\ell]$ plane and
    (b)~the $[hk0]$ plane in reciprocal space.
    The dispersive bands are twofold degenerate, while the zero-energy bands are fourfold degenerate.}
\label{fig:bands_CLTE}
\end{figure}

In this appendix, we provide a brief overview of the classical low-temperature expansion and its application to the $A_2\oplus T_1\oplus T_2$ line for the case $J_{z\pm}=0$. The classical low-temperature expansion is an analytical method that examines fluctuations around a ground state configuration by studying an effective low-temperature Hamiltonian. This effective Hamiltonian is obtained by expressing the classical spin vectors as small fluctuations about the ground state, namely
\begin{equation}
    \mathbf{S}_{i}\simeq \left(\delta n^{\tilde x}_{i},\delta n^{\tilde y}_{i}, S\left(1-\frac{(\delta n^{\tilde x}_{i})^2}{2S^2}-\frac{(\delta n^{\tilde y}_{i})^2}{2S^2}\right)\right)\label{eq:CLTE},
\end{equation}
where the basis $\{\tilde x,\tilde y, \tilde z\}$ is chosen such that the $\tilde{z}_i$ axis aligns with the selected ground state spin configuration at a given pyrochlore lattice site $i$. Substituting Eq.~\eqref{eq:CLTE} into the Hamiltonian given by Eq.~\eqref{eq:general_bilinear_hamiltonian} under the assumption that $\delta n^\alpha \ll 1$, we obtain an $8\times 8$ Hamiltonian acting on the spin fluctuations $\delta n^\alpha$, denoted $\mathcal H^{(2)}$, which can be diagonalized. For more details on this method, we refer to Ref.~\cite{Noculak_HDM_2023}.

As discussed in the main text, the application of a classical low-temperature expansion to the $A_2\oplus T_1\oplus T_2$ line for the case $J_{z\pm}=0$ leads to the observation of four zero-energy flat bands in the spectrum of $\mathcal H^{(2)}$, see Fig.~\ref{fig:bands_CLTE}.
At the present order of the expansion, the zero-energy bands can be understood to describe higher-order modes.
If these modes are considered to be quartic, the equipartition theorem predicts a low-temperature specific heat of $3/4$, consistent with the classical Monte-Carlo results.

%%%%%%%%%%%%%%%%%%%%%%%%%%%%%%%%%%%%%%%%%%%%%%%%%%%%%%%%%%%%%%%%%%%%%%%
\section{Specific heat in the AIAO phase of the Kramers Hamiltonian }
\label{appendix:C_v}
%%%%%%%%%%%%%%%%%%%%%%%%%%%%%%%%%%%%%%%%%%%%%%%%%%%%%%%%%%%%%%%%%%%%%%%

\begin{figure}[tb!]
    \centering
    \begin{overpic}[width=0.8\columnwidth]{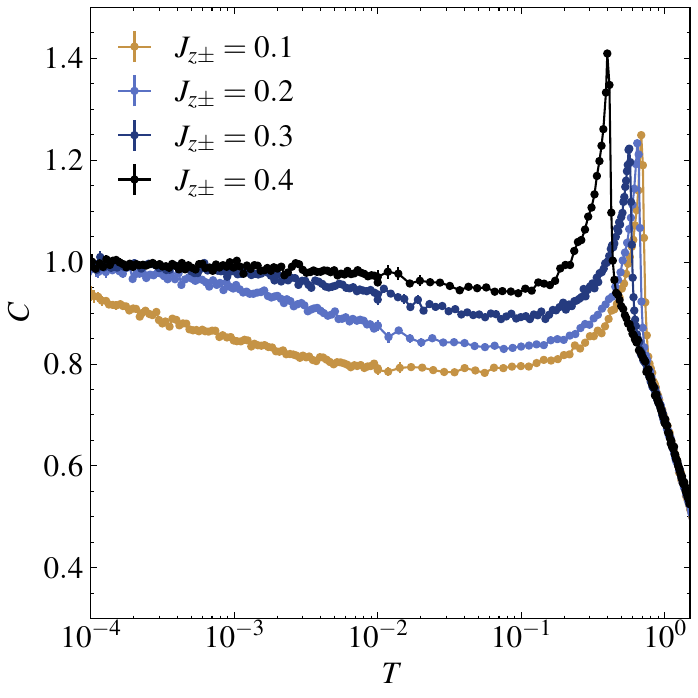}
    \end{overpic}
    \caption{%
    Specific heat as function of temperature in a linear-log plot
    from classical Monte-Carlo simulations on an $L=10$ lattice
    for different values of $J_{z\pm}$, indicating that $C(T\to0) =1$ in the $A_2$ phase of the Kramers Hamiltonian for $J_{z\pm} \neq 0$.}
    \label{fig:low-T-specific-heat}
\end{figure}

In this appendix, we show, using classical Monte-Carlo simulations, that the specific heat approaches one in the low-temperature limit of the $A_2$ phase in the Kramers case for $J_{z\pm} \in (0,0.5)$.
Upon cooling below $\Tc$, the specific heat initially drops below one, as illustrated in Fig.~\ref{fig:A2_CMC}(a). However, in the Kramers case for $J_{z\pm} \neq 0$, the classical low-temperature expansion now features eight dispersive bands at the quadratic order for the AIAO phase, in contrast to the non-Kramers case for $J_{z\pm} = 0$, thus predicting a specific heat value of one at low temperatures. Indeed, pushing the classical Monte-Carlo simulations down to temperatures as small as $T=10^{-4}$, the upturn of the specific heat toward $C=1$ becomes visible, see Fig.~\ref{fig:low-T-specific-heat}. All curves for $J_{z\pm} \geq 0.2$ converge at the predicted value $C(T\to 0) = 1$ at the lowest temperatures. The fact that the curve for $J_{z\pm}=0.1$ requires even smaller temperatures in order to observe convergence can be explained by the smaller bandwidth of the lowest-energy modes in the classical low-temperature expansion.
The same argument explains the behavior of the curves for different $J_{z\pm}$. Indeed, the band dispersion of the lowest mode monotonically increases with increasing $J_{z\pm}$ (not shown). Hence, the temperature regime for which the specific heat is one also increases with $J_{z\pm}$, as evident in Fig.~\ref{fig:low-T-specific-heat}.

%%%%%%%%%%%%%%%%%%%%%%%%%%%%%%%%%%%%%%%%%%%%%%%%%%%%%%%%%%%%%%%%%%%%%%%
\section{Temperature evolution of static structure factor}
\label{appendix:Structure_factors}
\begin{figure}[tb!]
    \centering
    \begin{overpic}[width=\columnwidth]{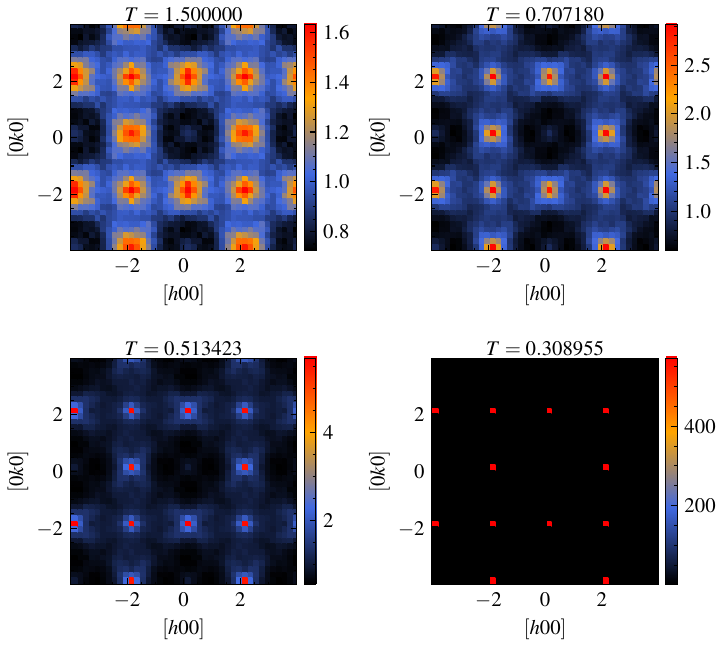}
    \put(75,81.5){\textcolor{white}{$\downarrow$}}
    \put(75,74){\textcolor{white}{$\uparrow$}}
     \put(85,60){\textcolor{white}{$\nwarrow$}}
     \put(79,60){\textcolor{white}{$\nearrow$}}
     \put(85,65){\textcolor{white}{$\swarrow$}}
     \put(79,65){\textcolor{white}{$\searrow$}}
    \put(4,90){(a)}
    \put(4,42){(c)}
    \put(55,90){(b)}
    \put(55,42){(d)}
    \end{overpic}
    \caption{%
    (a)~Static structure factor in the $[hk0]$ plane obtained via classical Monte-Carlo simulations for $J_{z\pm}=0.4$ and temperature $T=1.5$, in the high-temperature paramagnetic regime.
    (b)~Same as (a), but for $T= 0.707$, in the classical spin liquid regime. Arrows indicate the two-fold and four-fold pinch-point features.
    (c)~Same as (a), but for $T=0.513$, slightly above the critical temperature $\Tc = 0.4214(2)$.
    (d)~Same as (a), but for $T=0.309$, in the $A_2$ symmetry-breaking phase.}
\label{fig:S(q)_temp_CMC}
\end{figure}

\begin{figure}[tb!]
    \centering
    \begin{overpic}[width=\columnwidth]{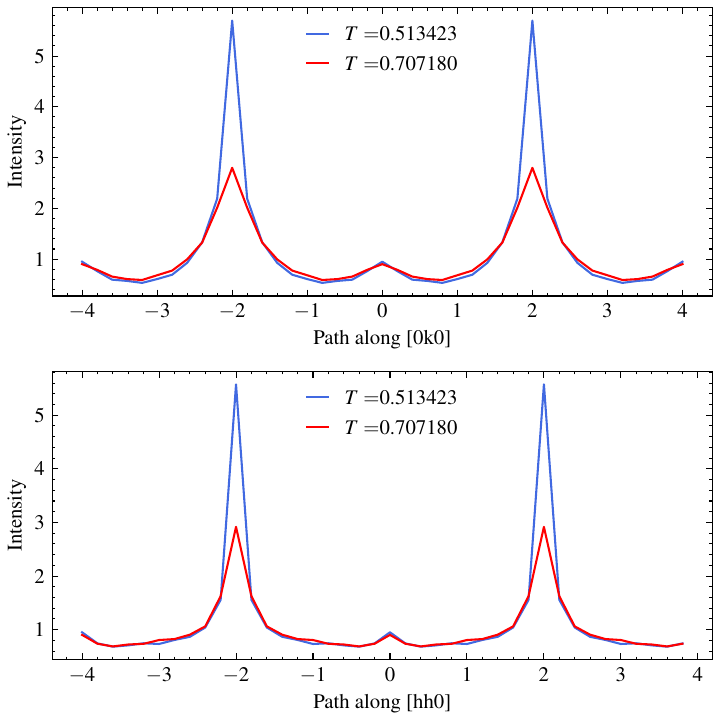}
    \put(2,99){(a)}
    \put(2,49){(b)}
    \end{overpic}
    \caption{%
    (a)~Cut of the structure factors shown in Figs.~\ref{fig:S(q)_temp_CMC}(b,c) along the $[0k0]$ direction in momentum space, corresponding to the low-intensity line of the two-fold pinch points at $[0\bar 20]$ and $[020]$. As the temperature decreases, the intensity shifts towards the Bragg-peak $\mathbf{q}$ points expected at low temperatures.
    (b)~Same as (a), but along the $[hh0]$ direction in momentum space, corresponding to the low-intensity line of the four-fold pinch points at $[220]$ and $[\bar 2 \bar 20]$.
    }
\label{fig:S(q)_cuts_temp_CMC}
\end{figure}

%%%%%%%%%%%%%%%%%%%%%%%%%%%%%%%%%%%%%%%%%%%%%%%%%%%%%%%%%%%%%%%%%%%%%%%
In this appendix, we provide the temperature evolution of the static structure factors obtained via classical Monte-Carlo simulations for the case $J_{z\pm}=0.4$.  As depicted in Fig.~\ref{fig:S(q)_temp_CMC} and discussed in the main text, the system goes from a high-temperature paramagnetic regime, where only broad features are observed, see Fig.~\ref{fig:S(q)_temp_CMC}(a), to a classical spin liquid phase where pinch-point features start to develop, as indicated by the white arrows in Fig.~\ref{fig:S(q)_temp_CMC}(b). The broadness of these pinch points is indicative of local violations of the energetically-imposed Gauss's law constraints describing the spin liquid phase~\cite{Conlon2010}. As the temperature is decreased, the pinch points become sharper. However, the proximity to the symmetry-breaking transition results in a shift of intensity towards the $\mathbf{q}$ vectors where Bragg peaks develop below the critical temperature $\Tc$, see Fig.~\ref{fig:S(q)_temp_CMC}(c) and Fig~\ref{fig:S(q)_cuts_temp_CMC}. Lastly, below the critical temperature $\Tc$, the system enters a symmetry-breaking phase, the antiferromagnetic $A_2$ phase, resulting in Bragg peaks in the static structure factor, see Fig.~\ref{fig:S(q)_temp_CMC}(d).
\\

%%%%%%%%%%%%%%%%%%%%%%%%%%%%%%%%%%%%%%%%%%%%%%%%%%%%%%%%%%%%%%%%%%%%%%%
\section{Neutron structure factors}
\label{appendix:Structure_factors}
%%%%%%%%%%%%%%%%%%%%%%%%%%%%%%%%%%%%%%%%%%%%%%%%%%%%%%%%%%%%%%%%%%%%%%%

\begin{figure}[tb!]
    \centering
    \begin{overpic}[width=0.8\columnwidth]{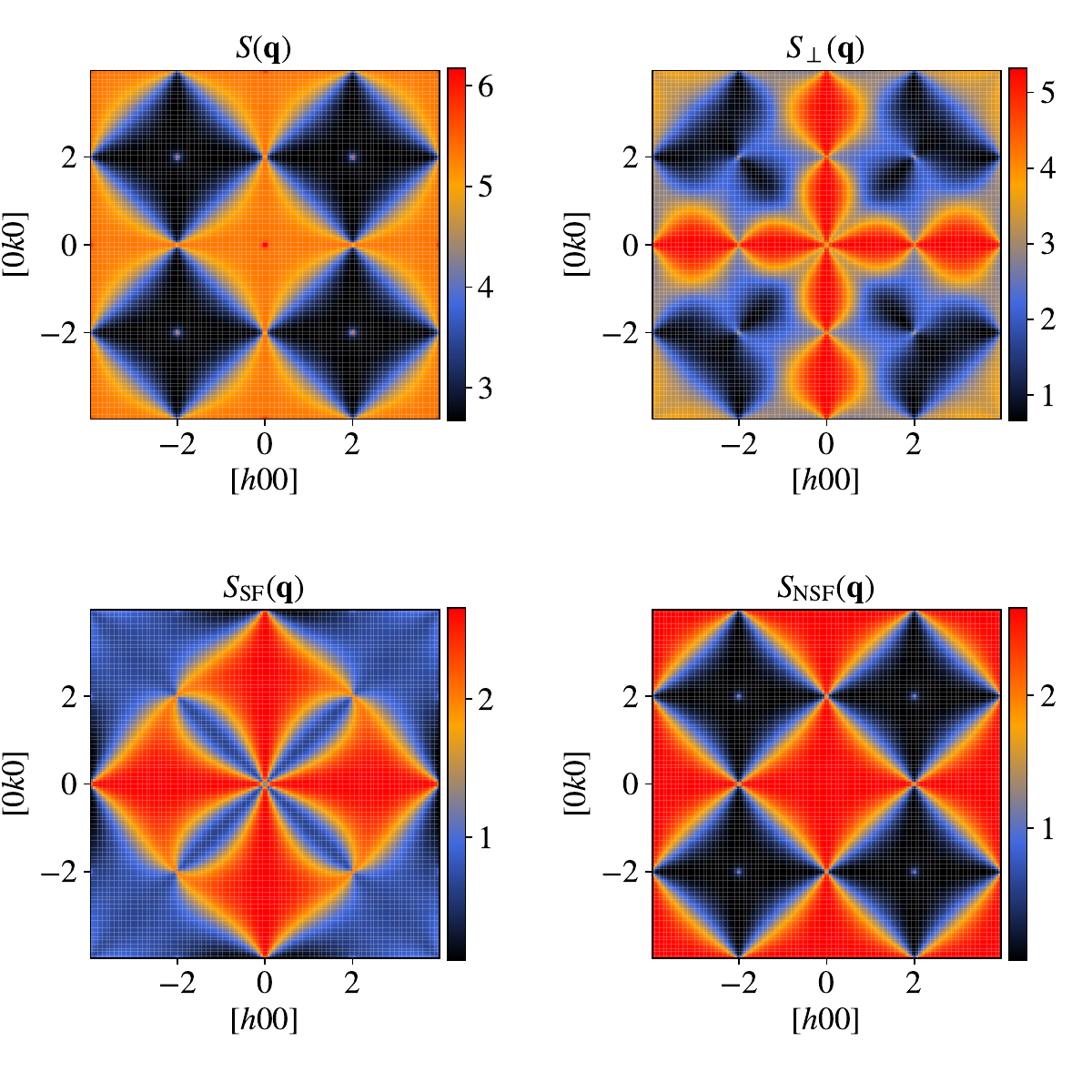}
    \put(4,97){(a)}
    \put(4,47){(c)}
    \put(55,97){(b)}
    \put(55,47){(d)}
    \end{overpic}
    \caption{%
    (a)~Static spin structure factor in the $[hk0]$ plane from SCGA in the low-temperature limit in the non-Kramers case $J_{z\pm}=0$ on the $A_2\oplus T_1 \oplus T_2$ line, describing the correlations in the rank-1 classical spin liquid. [Same as Fig.~\ref{fig:non-kramers}(a) in the main text.]
    (b)~Same as~(a), but showing the unpolarized neutron structure factor.
    (c)~Same as~(a), but showing the polarized neutron structure factor in the spin-flip channel.
    (d)~Same as~(a), but showing the polarized neutron structure factor in the non-spin-flip channel.
    }
\label{fig:S(q)_Jzpm0}
\end{figure}

\begin{figure}[tb!]
    \centering
    \begin{overpic}[width=0.8\columnwidth]{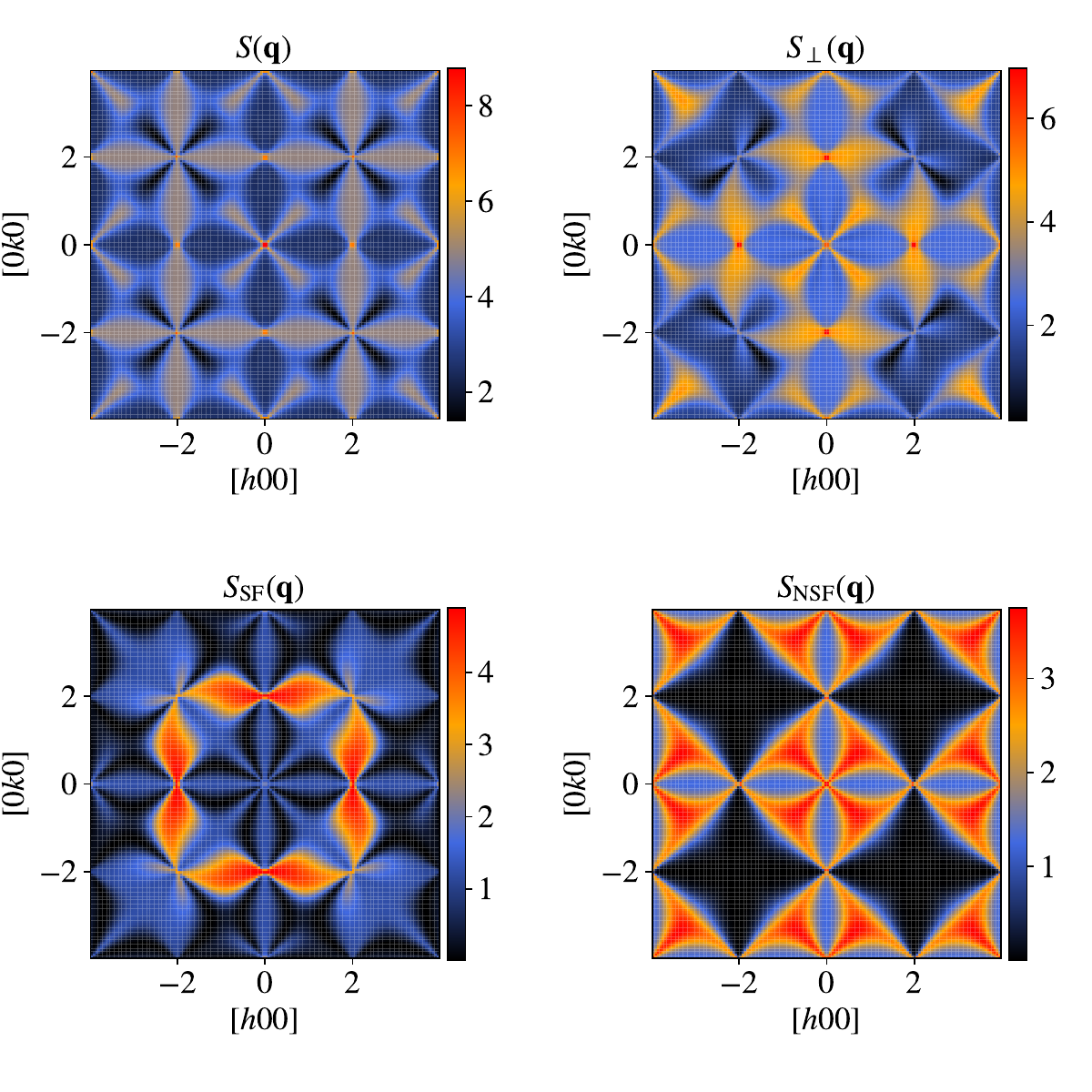}
    \put(4,97){(a)}
    \put(4,47){(c)}
    \put(55,97){(b)}
    \put(55,47){(d)}
    \end{overpic}
    \caption{%
    Same as Fig.~\ref{fig:S(q)_Jzpm0}, but in the Kramers case for a representative small value of $J_{z\pm} = 0.4$, describing the correlations in the rank-2 classical spin liquid.}
\label{fig:S(q)_Jzpm04}
\end{figure}

\begin{figure}[tb!]
    \centering
    \begin{overpic}[width=0.8\columnwidth]{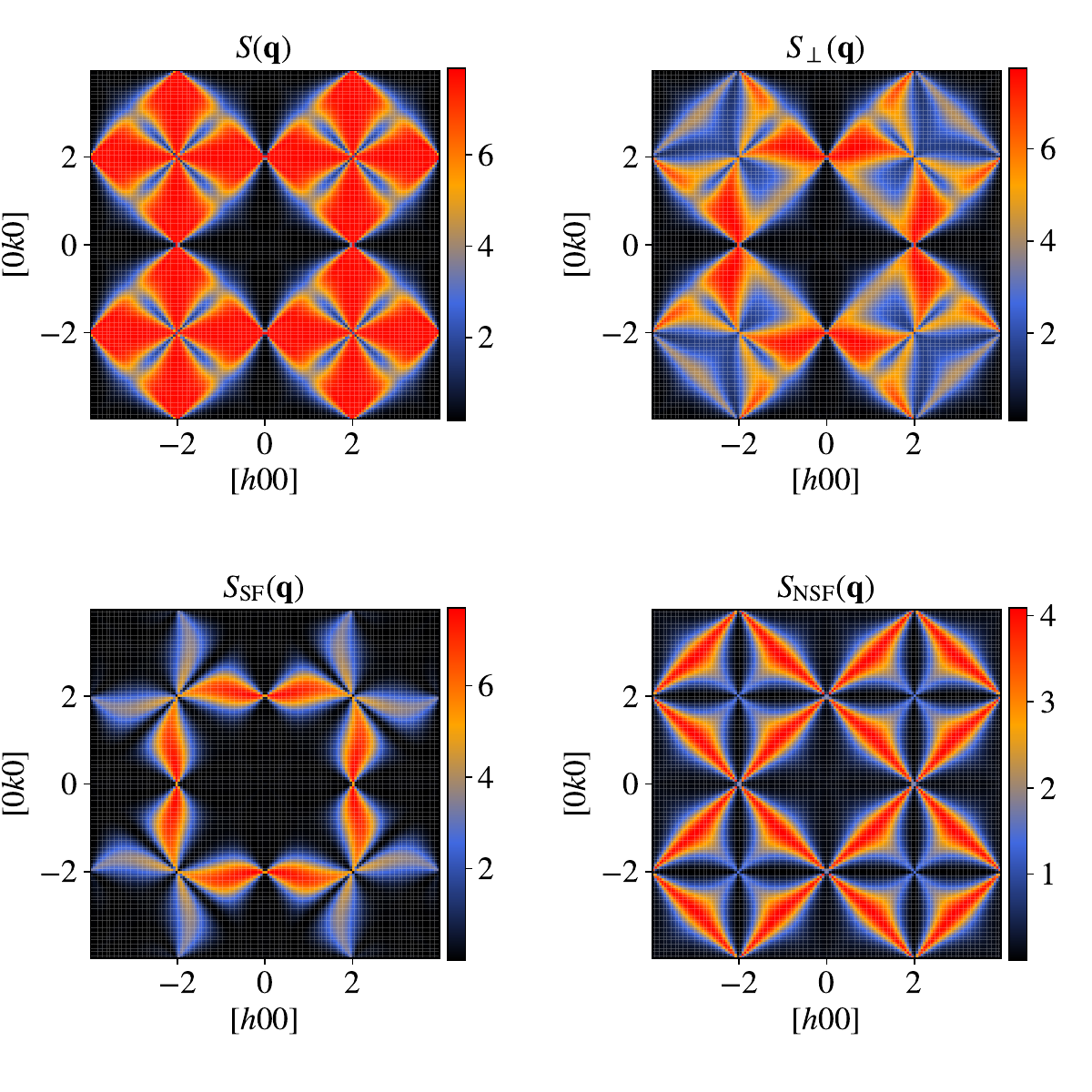}
    \put(4,97){(a)}
    \put(4,47){(c)}
    \put(55,97){(b)}
    \put(55,47){(d)}
    \end{overpic}
    \caption{Same as Fig.~\ref{fig:S(q)_Jzpm0}, but in the Kramers case for a representative large value of $J_{z\pm} = 1$, describing the correlations in the rank-2 classical spin liquid.}
\label{fig:S(q)_Jzpm1}
\end{figure}

\begin{figure}[tb!]
    \centering
    \begin{overpic}[width=0.8\columnwidth]{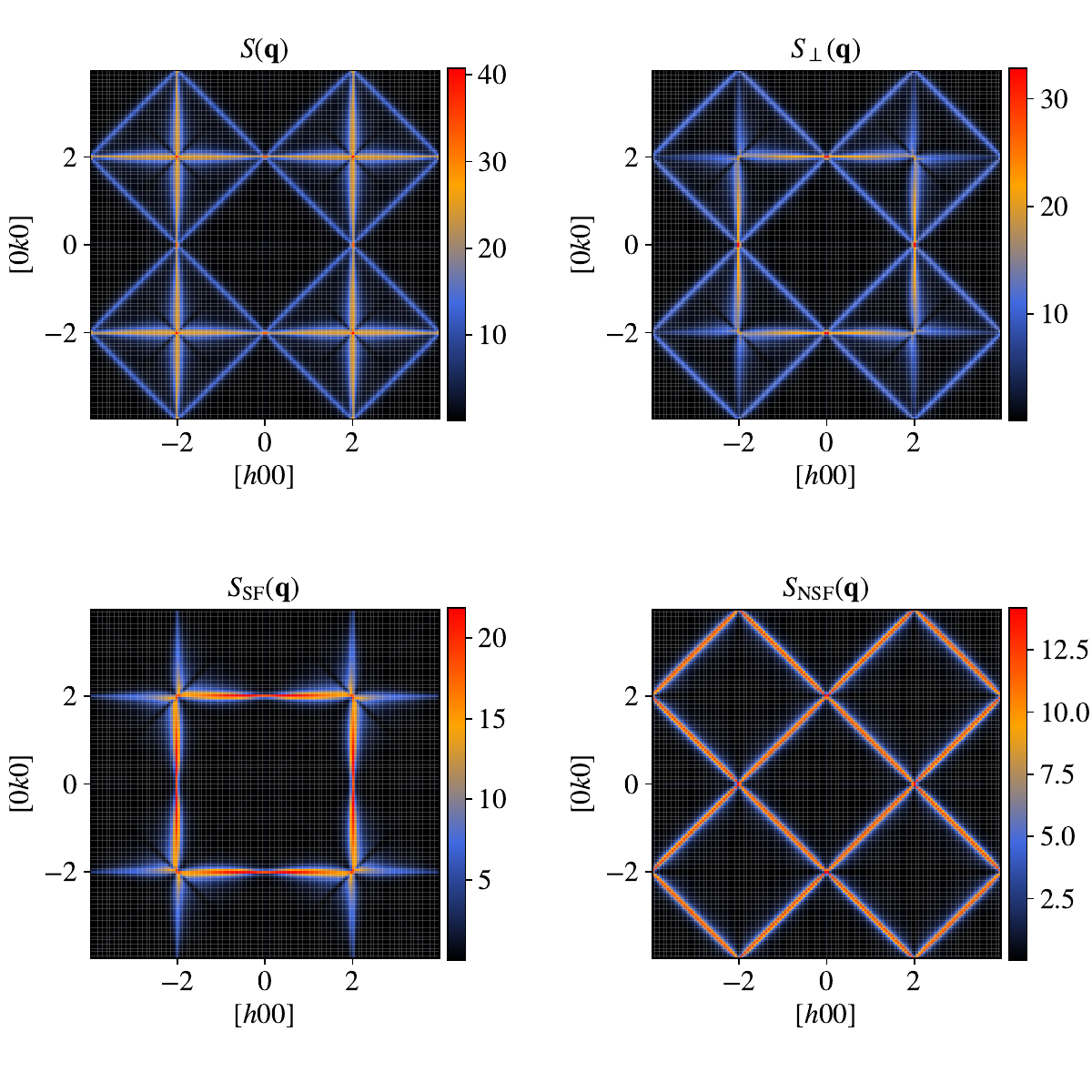}
    \put(4,97){(a)}
    \put(4,47){(c)}
    \put(55,97){(b)}
    \put(55,47){(d)}
    \end{overpic}
     \caption{Same as Fig.~\ref{fig:S(q)_Jzpm0}, but in the Kramers case for a representative value of $J_{z\pm} =1$ and $J_{\pm} = -0.625$ on the $T_1 \oplus T_2$ plane, describing the correlations in the spin nematic phase.}
\label{fig:S(q)_Jzpm_nematic}
\end{figure}

In this appendix, we provide SCGA neutron structure factors for both polarized and unpolarized cases in the classical spin liquid and spin nematic phases, to facilitate comparison with future experiments.
We define the unpolarized neutron structure factor as
\begin{eqnarray}
    S_\perp (\mathbf{q})=\sum_{\alpha\beta}\sum_{\mu\nu}\left(\delta_{\alpha \beta} - \frac{q^{\alpha} q^\beta }{q^2}\right)\langle S^\alpha_\mu(-\mathbf{q})S^\beta_\nu(\mathbf{q})\rangle, \label{eq:neutron_Sq}
\end{eqnarray}
and the non-spin-flip and spin-flip neutron structure factors as
\begin{align}
    S_\perp^{\rm NSF} (\mathbf{q})&=\sum_{\alpha\beta}\sum_{\mu\nu} z^{\alpha}_s z^{\beta}_s \langle S^\alpha_\mu(-\mathbf{q})S^\beta_\nu(\mathbf{q})\rangle, \label{eq:neutron_Sq_polarized_NSF}
\end{align}
and
\begin{align}
     S_\perp^{\rm SF} (\mathbf{q})&=S_\perp (\mathbf{q})-S_\perp^{\rm NSF} (\mathbf{q}),
    \label{eq:neutron_Sq_polarized_SF}
\end{align}
respectively.
In the above equations, the indices $\alpha$ and $\beta$ label the spin components, while $\mu$ and $\nu$ are sublattice indices, and $\bm z_s$ corresponds to the polarization direction of the incident neutron.
We consider structure factors in the $[hk0]$ plane in reciprocal space, as these are the ones where fourfold pinch points can be observed. Figure~\ref{fig:S(q)_Jzpm0} shows the spin structure factor together with the neutron structure factors for the rank-1 classical spin liquid in the non-Kramers case for $J_{z\pm} = 0$. For comparison, the spin and neutron structure factors in the rank-2 classical spin liquid in the Kramers case for two representative values of $J_{z\pm} = 0.4$ and $J_{z\pm} = 1.0$ on the $A_2\oplus T_1 \oplus T_2$ line are shown in Figs.~\ref{fig:S(q)_Jzpm04} and \ref{fig:S(q)_Jzpm1}.
Finally, Fig.~\ref{fig:S(q)_Jzpm_nematic} shows the spin and neutron structure factors for the spin nematic phase in the non-Kramers case for a representative value of $J_{z\pm} = 1$ and $J_{\pm} = -0.625$ on the $T_1 \oplus T_2$ plane.

%

%%%%%%%%%%%%%%%%%%%%%%%%%%%%%%%%%%%%%%%%%%%%%%%%%%%%%%%%%%%%%%%%%%%%%%%
\begin{acknowledgments}

We thank Michel J.\ P.\ Gingras, Han Yan, Owen Benton, Ludovic Jaubert, Johannes Reuther, Yasir Iqbal, Kristian Chung, and Matthias Vojta for insightful discussions.
This work has been supported by the Deutsche Forschungsgemeinschaft (DFG) through SFB 1143 (A07, Project No.\ 247310070), the W\"urzburg-Dresden Cluster of Excellence \textit{ct.qmat} (EXC 2147, Project No.\ 390858490), and the Emmy Noether program (JA2306/4-1, Project No.\ 411750675).
DLG acknowledges financial support from the DFG through the Hallwachs-R\"ontgen Postdoc Program of \textit{ct.qmat} (EXC 2147, Project No.\ 390858490).
The authors gratefully acknowledge the computing time made available to them on the high-performance computer at the NHR Center of TU Dresden. This center is jointly supported by the German Federal Ministry of Education and Research and the state governments participating in the NHR~\cite{nhr-alliance}.

\end{acknowledgments}
%%%%%%%%%%%%%%%%%%%%%%%%%%%%%%%%%%%%%%%%%%%%%%%%%%%%%%%%%%%%%%%%%%%%%%%

%%%%%%%%%%%%%%%%%%%%%%%%%%%%%%%%%%%%%%%%%%%%%%%%%%%%%%%%%%%%%%%%%%%%%%%
% BIBLIOGRAPHY: FOR USE WITH BIBTEX
%%%%%%%%%%%%%%%%%%%%%%%%%%%%%%%%%%%%%%%%%%%%%%%%%%%%%%%%%%%%%%%%%%%%%%%
\bibliographystyle{longapsrev4-2}
\bibliography{pyrochlore-nematic}
%%%%%%%%%%%%%%%%%%%%%%%%%%%%%%%%%%%%%%%%%%%%%%%%%%%%%%%%%%%%%%%%%%%%%%%

\end{document}